# BASIC CONCEPTS FOR A FUNDAMENTAL AETHER THEORY[1]

## Joseph Levy

4 square Anatole France, 91250 St Germain-lès-Corbeil, France
E-mail: levy.joseph@orange.fr

**ABSTRACT**

In the light of recent experimental and theoretical data, we go back to the studies tackled in previous publications [1] and develop some of their consequences. Some of their main aspects will be studied in further detail. Yet this text remains self-sufficient. The questions asked following these studies will be answered. The consistency of these developments in addition to the experimental results, enable to strongly support the existence of a preferred aether frame and of the anisotropy of the one-way speed of light in the Earth frame. The theory demonstrates that the apparent invariance of the speed of light results from the systematic measurement distortions entailed by length contraction, clock retardation and the synchronization procedures with light signals or by slow clock transport. Contrary to what is often believed, these two methods have been demonstrated to be equivalent by several authors [1]. The compatibility of the relativity principle with the existence of a preferred aether frame and with mass-energy conservation is discussed and the relation existing between the aether and inertial mass is investigated. The experimental space-time transformations connect co-ordinates altered by the systematic measurement distortions. Once these distortions are corrected, the hidden variables they conceal are disclosed. The theory sheds light on several points of physics which had not found a satisfactory explanation before. (Further important comments will be made in ref [1d]).

---

[1] Version supplemented by further explanations published in "Ether space-time and cosmology", volume 1, Michael C. Duffy and Joseph Levy Editors.

At the beginning of the twentieth century, the invariance of the speed of light and the application of the relativity principle in the physical world were regarded by almost all the physicists' community as undeniable assumptions. These concepts have been called into question only recently and we have the advantage of being informed of the debates raised around these subjects and of the new experimental results.
We should pay tribute to our predecessors who, despite their ignorance of these data, enabled physics to make noticeable progress.

**Foreword**
The study tackled in this text is partly based on ideas expressed in previous works [1a, 1b, 1c] but, in the light of new data, some main aspects of aether theory which had been treated in these works will be revisited and supplemented. The questions asked to the author will be answered.

Notice that, although it makes reference to papers already published, this text remains self sufficient. As we shall see, the theory developed here differs in various aspects from conventional relativity, and calls into question the approaches that deny the existence of the aether, or which suppose that there is no aether frame with specific properties that can be identified. It explains why the assumptions of Lorentz [2] prove today far better justified than in the past and it sheds light on some aspects of physics, never correctly explained before.

Yet, the transformations of Lorentz-Poincaré are exactly applicable exclusively when one of the reference frames they connect is the aether frame[*], in which the space and time co-ordinates are not altered by measurement distortions [1d], and it was necessary to search for a set of transformations applicable between any pair of 'inertial frames'. Contrary to what conventional relativity implies, we show that the transformations, which are derived on the basis of experimental data, result from the unavoidable measurement distortions due to length contraction and clock retardation, and from the arbitrary clock synchronization procedures which affect the Galilean coordinates. Of course the magnitude of the distortions varies as a function of the absolute speed of the frames under consideration, in contradiction with the relativity principle. Our approach differentiates completely from those of the classical authors (Lorentz, Larmor, Voigt, among others).

---

[*] In ref [1a] and [1b], we have applied the term Lorentz-Poincaré transformations to the transformations which assume the same mathematical form as the conventional transformations. In fact, as we shall see in ref [1d], the term should be reserved more specifically for the transformations which connect any 'inertial system' to the fundamental frame, in which the space and time co-ordinates are not altered by measurement distortions (which is not the case in the usual applications).

Notice that, the frames associated to bodies which are not submitted to external forces other than the interaction with the aether are called 'inertial' in this text, a term sanctioned by use. Yet, this designation is only appropriate when this interaction is weak. In this case, these frames can be regarded as almost inertial. But, as the interaction increases, the inertial character is lost.

**I. Introduction**

In the days when the result of Michelson's experiment was disclosed, Lorentz tried to explain the process without dismissing the concept of an absolute aether frame by means of the following assumptions [2]:

- existence of a privileged inertial frame attached to the aether, speed of light isotropic and of magnitude $C$ in this aether frame and different from $C$ in all other 'inertial frames', length contraction and clock retardation. The time was considered real in the aether frame and fictitious in all other frames, which implied that the measurement of the time in those frames was distorted.

(Note that the approach of Lorentz required equality of the two way transit time of light along the two arms of the interferometer. The little difference which was observed was considered negligible and dependent on the inaccuracy of the measurement. (Although it was probably due to the fact that the experiment was not made in vacuum, see later) The modern versions of Michelson's experiment in vacuum [3] have considerably reduced this inequality (Jaseja, Joos, Brillet and Hall, Müller et al)).

We must realize that when Lorentz formulated his assumptions, although at first many physicists were convinced of the existence of an absolute aether frame, there were no decisive arguments in favour of such a privileged frame. In addition, most of them believed that the relativity principle could be applied without any restriction in the physical world and, progressively, they began to suspect that the absolute aether frame was not compatible with the relativity principle. So, following the example of Einstein, they were led to abandon the aether frame in favour of the relativity principle.

Moreover, all the measurements of the speed of light gave $C$ in any direction of space, and in frames having no special properties (different from a privileged inertial frame [4]). The measuring process of this speed was generally considered indisputable and, therefore, was not called into question (see later and [1] or [31]). All these considerations looked to be in disagreement with the Lorentz assumptions.

(Note that Poincaré tried to reconcile the application of the relativity principle with the Lorentz assumptions [5] but, as we shall see, when the measurement distortions mentioned above are corrected, the two concepts prove in fact incompatible [1d]). The way was therefore paved for the approach of Einstein which assumed absolute invariance of the speed of light in all 'inertial frames' and in all directions of space and equivalence of all these frames for the description of the laws of physics [6]. The said assumptions implied reciprocity of observations (from one 'inertial frame' to another), relativity of time and relativity of simultaneity. (Contrary to Lorentz's approach, the theory did not assume the existence of a privileged time, all the times measured in the different 'inertial frames' had an equivalent status). Michelson's experiment was easily explained as follows: the two arms of the interferometer being equal, and the speed of light being isotropic, no fringe shift could occur. (The little fringe shift observed was considered negligible and resulting from experimental error). Note that, for Einstein, length contraction was assumed to be reciprocal and observational and, therefore, was nonexistent for an observer standing in the same frame as the arm under consideration.

Being convinced of the reliability of the experimental data available at that time, most physicists showed agreement with Einstein who proposed an approach in accordance with these results.

But, today, a number of theoretical and experimental arguments argue in favour of the existence of an absolute aether frame and of the anisotropy of the one-way speed of light in the Earth frame (see later).

There is no doubt that the relativity principle would exactly apply if the frames associated to bodies not submitted to physical influences other than the aether were perfectly inertial. But insofar as the bodies in motion are submitted to an aether drift, whose magnitude varies as a function of their speed with respect to the fundamental frame, the term 'inertial', applied to these frames, is inappropriate.

The anisotropy of the one-way speed of light gives a physical basis to the explanations of Larmor [7], Lorentz [2] and several prominent modern physicists who assume length contraction [8-17]. It challenges the other theories: Ritz theory [18], (ballistic) Einstein theory [6], Stokes theory [19] (completely dragged aether). As we shall see, the one-way velocity of light is erroneously found to be constant, because the measurements of the lengths and of the time are altered by systematic distortions due to length contraction, clock retardation and unreliable clock synchronization (Einstein-Poincaré procedure or slow clock transport).

Note that, insofar as the velocity of light is not exactly constant, the question of whether photons have a rest mass arises. Although the topic cannot be studied in detail here, we should note that, following Einstein and de Broglie, several authors in the past have discussed the possibility of a non-zero photon mass. More recently, Van Flandern and Vigier [20], making reference to the Walker-Dual experiment, according to which electrodynamic fields propagate faster than light (a result compatible with causality in Lorentz's approach), noticed that the experiment, if confirmed, suggests a non-zero photon mass. Among the other recent publications [21, 22], Prokopec et al suggest that "the magnetic fields that seem to permeate the Cosmos, might have arisen if photons possessed mass during the Universe early moments of expansion". One may notice that a non-invariant one-way velocity of light implies that one can envisage a rest frame for the photon, a fact which was not possible with special relativity. This result removes an obstacle to the existence of a rest energy for this particle, and therefore of a rest mass. Of course it is not sufficient by itself to conclude; only the experiment can do this. The difficulty comes from the smallness of such a hypothetic mass: the most precise direct bounds yield $m \leq 10^{-47} g$, [22].

We propose here:
1. To give different arguments which call into question the conventional approach of relativity, and permit to show:
-the necessity of a fundamental inertial frame,
-that relativity of simultaneity is only apparent and results from a confusion made between the instantaneous signals, and the light issued from them.
2. To demonstrate that: -Assuming the recent arguments in favour of the anisotropy of the one-way speed of light, length contraction is no longer an ad hoc hypothesis, but a necessary cause of the Michelson result.
-The apparent invariance of the two way speed of light results from the systematic measurement distortions caused by length contraction and clock retardation.
3. To derive a set of space-time transformations based on the Lorentz assumptions which apply between any pair of 'inertial frames'. After correction of the measurement distortions, the hidden variables they conceal are disclosed.
4. To point out some consequences of the theory:
    a. The classical law of variation of mass with speed is completely exact, only if the mass of a body in motion is compared to its mass in the fundamental frame.

b. Insofar as an aether drift exists, the relativity principle cannot exactly apply in the physical world [1d].
c. The compatibility of the relativity principle with mass-energy conservation is checked and the origin of mass is discussed.

**II. Let us first bear in mind two decisive arguments in support of the existence of an aether frame and which call into question the application of the relativity principle in the physical world.**

The experimental arguments will be briefly presented in section II.3.

**II.1 Argument relative to the lifetime of the muon [23]**

The $\mu^+$ muons were discovered in 1935 in the cosmic radiation. They are unstable particles whose decay gives classically a positron, a neutrino and an anti-neutrino ($\mu^+ \rightarrow e^+ + \nu^+ + \nu^-$). The half-life of these muons when they are at rest in the Earth frame, $t$, is about $2.2 \ 10^{-6}$ sec. In the high atmosphere, the muons move at a speed close to the speed of light. Their mean free path should therefore be approximately $L = Ct = 660$ m. Nevertheless the measurements carried out for $L$ give a much greater value, of the order of several kilometres, which corresponds to a half-life $T$ much longer.

According to special relativity, the proper lifetime of a particle is independent of its speed with respect to another frame. Therefore, the half-life of a high energy muon measured by an observer at rest with respect to it should also be $2.2 \ 10^{-6}$ sec. This statement is nevertheless questionable. The reasons which justify its challenge rest on the three following propositions:

The real relative speed between two bodies A and B, receding from one another uniformly along the same line is invariant. It is the same for A and for B.

The real relative distance does not depend either on which one measures it.

Consequently, the real time needed by the bodies to recede from each other from distance zero to distance $l$ must be the same for A and for B.

We can therefore conclude that the proper half-life of the muon at a speed close to the speed of light must also be $T$. These considerations lead us to infer that there is a difference between rest and motion, which implies that motion possesses an absolute character. In other words, the "rest frame" and the "moving frame" are not equivalent and we are led to recognize the existence of a privileged aether frame designated as "Cosmic Substratum".

Apparent coordinates.

The half-life of the muon at high speed is *T*. But clocks moving with it (at rest with respect to it) would display the reading

$$t = T\sqrt{1 - v^2/C^2}$$

So, *t* is not the real half-life, but the clock reading due to clock retardation. (This result is only approximate, because as we shall see, the velocity of the Earth frame with respect to the Cosmic Substratum is not null, but this velocity is very small compared to the speed of high-energy muons which approximates the speed of light).

## II.2 Poincaré's relativity principle revisited

This question has been treated in ref [1]. We propose to revisit the subject and to add some comments in response to the questions asked.

According to Poincaré's relativity principle, it would be impossible by means of an experiment internal to a given 'inertial frame' S to know whether this frame is at rest or in motion with respect to the aether frame. This statement is questionable [23] as the following demonstration will show: consider two vehicles moving uniformly in opposite directions along a straight line. At the instant $t_0$, they meet at a point O of frame S and then continue on their way symmetrically, with identical speed *v*, towards two points A and B placed at equal distances from point O. At the instant they meet, the clocks inside the vehicles are set to $t_0$. According to the relativity principle, the clocks should display the same time when they reach points A and B. If this were not the case, this would represent a criterion capable of determining if frame S is in motion or at rest with respect to the aether frame. But, insofar as the vehicles do not have the same speed with respect to the aether frame, the slowing down of their clocks with respect to the clocks at rest in this privileged frame will be different and, therefore, their indications will be different. Logically, Poincaré's relativity principle would exactly apply only if the frames associated to moving bodies not submitted to external physical forces could be regarded as really inertial. But this opinion cannot be true in all generality, since it ignores the role of the aether. The inertial character must be viewed as an approximation only valid when the aether drift is weak.

Therefore the exact applicability of the relativity principle proves incompatible with the existence of an aether frame in a state of absolute rest. Since there are today a number of theoretical (see for example section II.1, and [1]) and experimental arguments (see later) lending support to the existence of an aether frame, the relativity principle cannot exactly apply.

*Discussion*

We note that in the experiment we have just mentioned, the speed *v* of the vehicles was exactly determined. If the speeds were measured with clocks placed in A, O and B, synchronized by means of the Einstein-Poincaré synchronization procedure, we can demonstrate that the clocks in the vehicles would display the same reading when they reach the points A and B [23]. But this result follows from the systematic error made in measuring the speeds when, using this synchronization procedure, one assumes the isotropy of the one-way speed of light. This systematic measurement error is needed in order for the laws of physics to assume an identical mathematical form in all 'inertial frames',

One cannot conclude that the relativity principle is a fundamental principle of physics if it depends on a measurement error.

## II.3 Experimental arguments

We have no means today to directly prove the existence of the aether, but its reality can be indirectly established by measuring the anisotropy of the one-way speed of light in the Earth frame. (Note that, as demonstrated by Anderson et al [24], all the recent experiments purporting to demonstrate the invariance of this velocity were based on erroneous ideas, because they assumed that the slow clock transport procedure allows exact synchronization. It is clear that this is not the case, (see ref [1]). On the contrary, a number of arguments lend support to the anisotropy of the one way speed of light. Although its direct evaluation encounters major difficulties, it can be deduced from the measurement of the absolute velocity of the solar system.

A first estimate of this velocity had already been made in 1968 by de Vaucouleurs and Peters [25], by measuring the anisotropy of the red shift of many distant galaxies. The experiment was repeated by Rubin in 1976 [25].Although modern authors consider that the method needs a small adjustment (see the entry relative to M. Allais in further references with comments) an estimate of the solar system absolute velocity can be obtained by measuring the anisotropy of the 2.7°K microwave background radiation uniformly distributed throughout the Universe. "An observer moving with velocity *v* relative to the microwave background radiation can detect a larger microwave flux in the forward direction $(+v)$ and a smaller microwave flux in the rearward direction $(-v)$. He can observe a violet shift in the forward direction $(+v)$ and a red shift in the rearward direction $(-v)$" (Wilhelm [26]). Using this method, a consensus was obtained by different experimenters ((Conklin (1969), Henry (1971), Smoot et al (1977), Gorenstein and Smoot (1981), Partridge (1988) [27])). Let us also quote the method of measurement based on the muon flux anisotropy (Monstein and Wesley (1996)) [28]. Wesley [29] and Wilhelm [26] give an assessment of all these experiments.

A verification of the absolute speed of the Earth frame was made by Roland De Witte in 1991. To that end, 5 MHz radio-frequency signals were sent in two opposite directions through two buried co-axial cables linking two caesium beam atomic clocks separated by 1.5 Km. Changes in propagation times were observed and were recorded over 178 days. De Witte interpreted the results as evidence of absolute motion [27].

More recently, Cahill and Kitto reinterpreted Michelson and Morley's experiments. They asserted that Michelson interferometers operating in gas mode are capable of revealing the Earth's absolute motion [27]. They analysed the old results from gas-mode Michelson interferometers experiments which always showed small but significant effects. In a first evaluation, the authors asserted that after correcting for the refractive index of the air, the Miller experiment gives a speed of $v=335\pm 57$ Km/sec. A more recent evaluation by Cahill yielded $420\pm 30$ Km/sec.

If confirmed, these results (from De Witte to Cahill) provide further weighty arguments in favour of the existence of a fundamental aether frame, in addition to the others arguments previously mentioned.

Marinov also attested having measured the absolute velocity of the solar system by means of different devices (coupled mirrors experiment and toothed wheels experiment [30]). According to Wesley [29], "the Marinov (1974, 1977a, 1980b) coupled mirrors experiment is one of the most brilliant and ingenious experiments of all time. It measures the very small quantity $v/C$, where $v$ is the absolute velocity of the observer, by using very clever stratagems". The coupled mirrors experiment enabled the author to assert that the solar system absolute velocity $v$ is of the order of $300 \pm 20$ km/sec and that the speed of light is $C-v$ in its direction of motion and $C+v$ in the opposite direction. (Notice that the orbital motion of the Earth around the sun is far slower (about 30 km/sec) and that the rotational motion at the latitude of the experiment was of the order of 0.5 km/sec).

All the above-mentioned experiments gave a result of the same order. They demonstrated the existence of a fundamental frame whose absolute speed is zero, but whose relative speed with respect to the Earth frame is of the order of 350 to 400 km/sec.

**III. Critical approach of the relativity of simultaneity.**

Let us briefly bear in mind the arguments of special relativity and succinctly reply. The question has also been studied in ref [32] in a different way. According to special relativity, two distant events, which are simultaneous for one observer, are not for another moving with respect to the first with rectilinear uniform motion.

In order to demonstrate this theorem, Einstein takes the classical example of the train and the two flashes of lightning [33]: two flashes strike at the two ends A and B of an embankment at the very instant when the middle O' of the train meets the middle O of the embankment (Einstein). By definition, the two flashes will be considered simultaneous with respect to the embankment if the light issuing from them meets the middle of the embankment at the same instant. Einstein adds that the definition is also valid for the train, but as the train travels towards point B, the light coming from B will reach the middle of the train before the light coming from A. Einstein concludes that two events simultaneous for the observer standing on the embankment are not simultaneous for the observer in the train (figure 1).

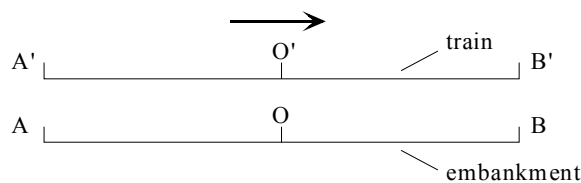

Figure1. At the initial instant, O and O' are coincident

As we have suggested in ref [34], Einstein's definition is only appropriate if the sources of light are firmly fixed to the reference frame in which the measurement is carried out. In the aforementioned example, if the light is emitted by two lamps attached to the embankment, the definition will be true for the embankment and not for the train and conversely. Indeed, Einstein himself recognizes that the light coming from B must cover a shorter path to reach O' than the light coming from A. But the definition is only valid if the light covers the same path in both directions.

(Note that the Earth also moves with respect to the train, and the embankment has no privileged status with respect to the train).

In order to correctly reason, we must bear in mind that simultaneous reception does not necessarily imply simultaneous emission and conversely. Hence, to accurately define simultaneity (if one assumes the invariance of the speed of light) we must specify the following points:

Two instantaneous events, occurring at two points A and B and emitting light in opposite directions toward a point O which is the middle of A B at the beginning of the experiment, can be considered simultaneous if the light issuing from A and B reaches point O at the same instant, provided that O remains fixed with respect to A and B all through the experiment.

We must add that this definition is exact, only if the speed of light is equal in both directions. In the fundamental aether theory the definition is considered exact in the aether frame and not in other frames, since in those frames the one-way speed of light is not isotropic.

We shall now propose another definition of simultaneity that is valid when point O' moves with respect to points A and B.

Let us reconsider to this end the example of the train and the embankment just seen (see fig 2). Contrary to ref [34] we shall not suppose a priori that the speed of light is identical in the opposite directions, so we shall designate the speed of light in the two reverse directions as $C_{AB}$ and $C_{BA}$. Let us place at A, O and B three clocks perfectly synchronous.

At the initial instant, O coincides with O'. At this very instant two signals are sent from A and B in opposite directions. When the signal coming from A reaches point O, point O' has moved towards B a distance equal to $\frac{v}{C_{AB}} \ell_0$, where $v$ is the speed of the train and $\ell_0 = AO$. When the signal has covered this latter distance, point O' has moved an additional distance:

$$\frac{v}{C_{AB}} \left( \frac{v}{C_{AB}} \ell_0 \right)$$

and so on. Thus, in order to reach point O' the signal must cover the distance

$$\ell_0 \left( 1 + \frac{v}{C_{AB}} + \frac{v^2}{C_{AB}^2} + \ldots + \frac{v^n}{C_{AB}^n} + \ldots \right) = \ell_0 \frac{C_{AB}}{C_{AB} - v}$$

and the time needed to cover this distance will be

$$t_{AO'} = \frac{\ell_0}{C_{AB} - v}$$

Now, in order to reach the middle of the train, the signal coming from B will cover a distance $x$ such that:

$$OO' = \frac{v}{C_{BA}} x = \ell_0 - x \quad \text{(See fig 2)}$$

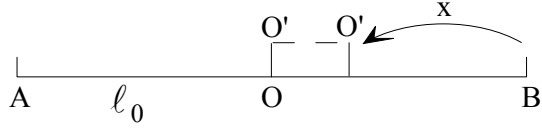

Figure 2. The light ray coming from B meets the middle of the train when the train has covered the distance OO'.

Therefore: 
$$x = \frac{\ell_0 C_{BA}}{C_{BA} + v}$$

and the time needed to cover the distance will be:
$$t_{BO'} = \frac{\ell_0}{C_{BA} + v}$$

*Hence, we can conclude that two instantaneous events occurring at A and B can be considered simultaneous if the light issuing from them reaches the middle of the train at two instants $t_{AO'}$ and $t_{BO'}$, such that*

$$t_{AO'} - t_{BO'} = \ell_0 \left( \frac{1}{C_{AB} - v} - \frac{1}{C_{BA} + v} \right)$$

<u>*Important remarks*</u>

1 - In reference [34] we assumed that the speed of light was isotropic such that $C_{AB} = C_{BA}$. This is true exclusively in the fundamental inertial frame. In this case we have.

$$t_{AO'} - t_{BO'} = \frac{2\ell_0 v}{C^2 \left(1 - v^2/C^2\right)}$$

where $C$ is the speed of light in the fundamental frame.

2 - Note that in this example, $t_{AO'}$ and $t_{BO'}$ are the real times given by the clocks attached to the privileged frame. As a result of clock retardation, the reading of the clocks in the train would be different. But this could not affect our reasoning and our conclusions regarding the absolute character of simultaneity. The same remark can be done about the length contraction affecting the train.

- Here is another example that will confirm this absolute character of simultaneity. It will, without doubt, convince the wavering reader.

Consider two rigid collinear rods AB and A'B' moving in opposite directions uniformly, along the same line. The rods are assumed to have identical length when they are in motion with relative speed *v*. They are firmly attached respectively to reference frames S and S' (see figure 3).

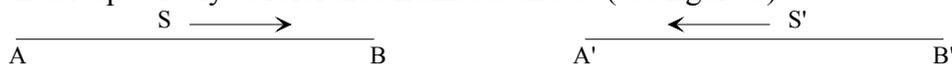

Figure3. An example of the absoluteness of simultaneity

In A and B are placed two identical clocks perfectly synchronous. Another pair of such clocks are placed in A' and B'. In order to reach A', clock A must cover a distance D=AA'. This is also the case for B in the direction of B' (since BB'=AA'=D). Therefore, for an observer in frame S, the encounters between A and A' and between B and B' will be simultaneous. But an observer in frame S' will draw the same conclusion. Hence the two observers conclude that both events are simultaneous. (Note that this does not imply that the clocks in frame S will display the same reading as those in frame S'. The conclusion concerns simultaneity and not the clock reading. We must distinguish between clock retardation and relativity of simultaneity).

We must be aware that an apparent relativity of simultaneity exists. It is inherent in the Lorentz-Poincaré transformations and the extended space-time transformations that will be studied later. But it is not essential and results from the systematic synchronism error entailed by the Einstein-Poincaré synchronization procedure (or by the slow clock transport method). These methods are not ideal but they are the most simple and the most often used.

After correction of the systematic measurement distortions inherent in these methods, the absolute character of simultaneity is found again.

A criterion of absolute simultaneity has already been given in a previous paper [32]: let two identical rubber balls, dropped from the same height, bounce on the two pans of a precision balance; if the central pointer of the beam does not move, we can consider that the balls have bounced at the same instant, and this is true for all observers, at rest or in motion with respect to the balance, whether they are accelerated or not. (Of course, a small correction would be necessary, since there could be a minute difference in the speed of propagation of the vibration along the two arms of the beam. But this is of no consequence since the correction would be identical for all the above-mentioned observers). Note also that the notion of four-dimensional space-time has also conventional character and once synchronism errors are corrected, the mixing of space and time disappears. This result will be demonstrated subsequently. (see also ref [35] and [32]).

**IV. Michelson's experiment**

Michelson's experiment in vacuum can be easily explained by means of Einstein's Special relativity: the two arms of the interferometer being considered equal, and the speed of light constant, the transit time of light in both directions must be identical. Now, assuming that the speed of light is anisotropic, this approach becomes questionable.

The explanation of the experiment based on Lorentz's assumptions is completely different. Let us suppose that at a moment in its journey, one of the arms of the interferometer moves along the *x*-axis of a co-ordinate system S attached to the cosmic substratum (aether). The other arm is aligned along the *y'*-axis perpendicular to the direction of motion of a coordinate system S' attached to the Earth platform (see fig. 4). We shall suppose that the experiment is carried out in vacuum. During a short instant, the motion of the Earth frame can be regarded as rectilinear and uniform: if this were not the case, we would be affected by the accelerations. Let us refer to the relative speed between S and S' as *v*.

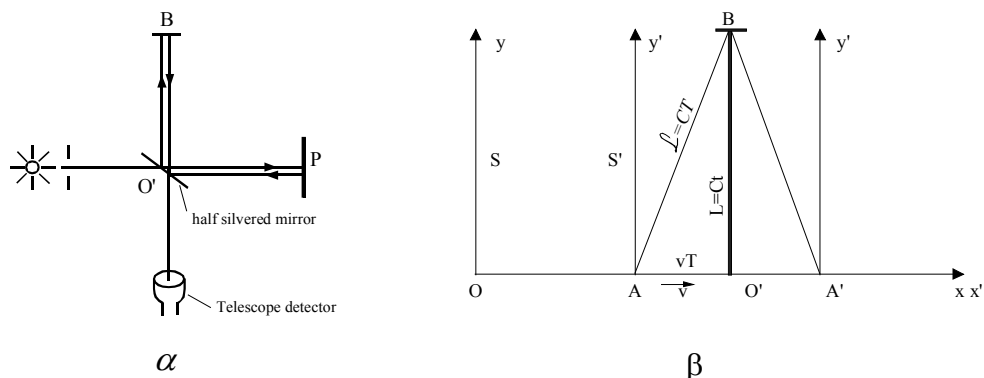

Figure 4.

*α*
Michelson's interferometer

*β*
The path of the light signal in the y' direction viewed by an observer attached to frame S: the signal starts from A, is reflected in B, and then comes back to A'

We shall first consider the arm perpendicular to the direction of motion.

From the point of view of an observer at rest in frame S', a light beam travelling back and forth along the arm O'B, covers a distance *2L*; but from the point of view of an observer attached to reference frame S, the beam starts from A, is reflected in B, and then comes back to A'. (Where AA' designates the distance covered by the interferometer during a cycle of the beam, see fig 4β). Since the speed of light is constant in the substratum, we have: AB = $\mathcal{L}$ = *CT*, where *T* is the time needed to cover the distance AB.

Now, from the classical (Galilean) viewpoint, the time separating two events is independent of the frame from which it is measured and, consequently, the speed of light must be lower than *C* in frame S'. Indeed, since $\mathcal{L} > L$ one cannot have at the same time: $\mathcal{L} = CT$ and $L = CT$. However, the measurement of the speed of light gives *C* in all 'inertial frames'. So there is a paradox. If we suppose that *C* = const, then, the time interval between the emission and the arrival of the beam must be different in the two frames, and we will have: $\mathcal{L} = CT$ and $L = Ct$. This is what special relativity asserts.

Nevertheless, if we give credit to the Lorentz assumptions, this result is questionable. The real value of the speed of light in frame S' is given by $(C^2 - v^2)T^2 = C'^2 T^2$. So that $C' = \sqrt{C^2 - v^2}$

How can we explain that we find *C* and not *C'*? In order to understand this, we must assume that the motion causes a slowing down of the clocks. Therefore, any measurement of the time in a frame moving with respect to the Cosmic Substratum will be different from the universal time. The relationship between the local (fictitious) time *t*, and the real (universal) time *T*, is easily obtained from fig.4:

$$T = \frac{t}{\sqrt{1 - v^2/C^2}} = \frac{L}{C\sqrt{1 - v^2/C^2}}$$

Therefore, if we assume the Lorentz postulates, there is no relativity of time but rather a slowing down of the clocks which are moving with respect to the aether frame. The real two way transit time of light along the arm O'B is:

$$2\frac{L}{C\sqrt{1 - v^2/C^2}} \tag{1}$$

Let us now consider the arm parallel to the direction of motion. The time needed by the light signal to travel back and forth along this arm was expected to be: $L\left(\frac{1}{C-v} + \frac{1}{C+v}\right) = 2\frac{L}{C(1 - v^2/C^2)}$ it was different from (1). Yet, the fringe shift observed when one changes the orientation of the interferometer was regarded as negligible in comparison with the shift expected. In order to explain this result, FitzGerald and Lorentz had the idea to postulate a contraction of the arm aligned along the *x*-axis, its length being equal to $\ell = L\sqrt{1-v^2/C^2}$. As a consequence, the two way transit time of light in vacuum, along both arms, proves identical, that is: $2\dfrac{L}{C\sqrt{1-v^2/C^2}}$. Note that, insofar as the anisotropy of the one way speed of light is proved, length contraction is no longer an ad hoc hypothesis, but rather a necessary cause of Michelson's result.

Now, we may wonder how Lorentz's theory can explain that the magnitude of the two-way speed of light along the *x'*-axis is always found to be *C*? since: $\dfrac{2L\sqrt{1-v^2/C^2}}{2L/\left(C\sqrt{1-v^2/C^2}\right)} = C\left(1-v^2/C^2\right)$

and not *C* ! There are two reasons for this:
1. The meter stick used to measure *L* is also contracted, so we cannot observe the contraction, and therefore we make a systematic error in measuring the arm. We find *L* in place of $L\sqrt{1-v^2/C^2}$

2. As we have seen from the study of the arm O'B, the clocks in the coordinate system S' slow down in such a way that the reading noticed by observer S' will be equal to the time noticed by S multiplied by $\sqrt{1-v^2/C^2}$

Finally, the *apparent* (experimental) average two-way speed of light will be
$$\dfrac{2L}{2L/C} = C \qquad (2)$$
(Note that it is this velocity which is generally identified with the one-way speed of light).

*Important remark*
It is essential to realize that *C* is not the real one-way speed of light in frame S'. It is well known that measuring exactly the one-way speed of light presents serious difficulties, since, to this end, we generally use clocks synchronized by means of Poincaré-Einstein's method, (or by slow clock transport). Indeed, suppose that we want to synchronize two clocks placed at two points O' and Q, aligned along the *x'*-axis of frame S'. According to the Einstein-Poincaré procedure, we send a light signal at time $t_0$ from O' to Q. After reflection, the signal returns to O'. The reading of the clock O' at this instant is $t_0 + \Delta t$. The clocks will be considered synchronous, if when the signal reaches clock Q the reading of this clock is $t_0 + \Delta t/2$. But, insofar as the one-way speed of light is anisotropic, the method introduces a systematic error. $\Delta t/2$ is not the one way transit time of light, it is the '*apparent*' average transit time (measured with the retarded clocks attached to frame S') that we shall call $\tau_{1app}$ (see section V.1.2). We see that the method only permits determination of the *apparent* average two-way speed of light along the *x'*-axis (measured with retarded clocks and contracted meter sticks).

As we have seen, and as we shall confirm in a more general way in the following chapters, it is this velocity which is always found to be constant and equal to *C* (see also [1 or 31]). (Notice that, contrary to a common belief, the method of slow clock transport does not provide more accurate results than the method of Einstein-Poincaré [1, 24]). Both methods are approximately equivalent.

**V. Transformations of space and time**

**V.1 Example of a light signal**

Many physicists consider that the theories which assume the existence of an aether frame and Einstein's relativity are equivalent. Yet, as we shall see, their physical meaning is completely different. It is not necessary here to derive Einstein's transformations, this has been done by different well-known techniques. On the other hand, the derivation of the transformations which assume the existence of the Lorentz aether and the variability of the one-way speed of light is not familiar. In order to determine them, we use a mathematical tool which makes reference to Zeno. The method discloses some hidden aspects of these transformations. In fact, the space-time transformations we shall derive take their usual mathematical form (Lorentz-Poincaré transformations) only when they connect the aether frame with any other reference frame not submitted to accelerations. In all other cases, they take another mathematical form.

(Yet, as we shall see in ref [1d], by using the apparent speeds instead of the real speeds, they can be converted into a set of equations which take the same mathematical form as the Lorentz-Poincaré transformations, but they differentiate from the Lorentz-Poincaré transformations and their meaning is quite different, because this mathematical form applies exclusively when the co-ordinates they connect are altered by the measurement distortions mentioned above, including in particular the arbitrary synchronization procedure which generates a synchronism discrepancy effect). And therefore, contrary to what relativity theory asserts, only the laws of physics relating distorted variables will be invariant and not the true laws, demonstrating the contingent character of the relativity principle. (It is clear that, as far as these transformations assume the variability of the one-way speed of light when this velocity is exactly measured, they cannot have the same meaning as the conventional transformations). We shall study successively the different cases.

**V.1.1** Consider first the co-ordinate systems S and S' mentioned above (fig 5). S is at rest in the Cosmic substratum, and S' moves away from S with rectilinear uniform motion along the *x-axis*. A long rod AB at rest in the co-ordinate system S' is aligned along the *x'*-axis. Let us name $\ell$ the length of the rod in S' and $v_0$ the relative speed between S and S'. At the initial instant $t_0$, the origins of the co-ordinate systems O and O' and the origin of the rod are coincident. At this instant, a light ray emanates from the common origin, and travels towards the end of the rod. (We must bear in mind that the speed of light is equal to $C$ in the substratum and different from $C$ in all other frames).

When the light ray has covered the distance $\ell$ in the substratum, the rod has covered the distance $\frac{v_0}{C}\ell$; as the ray has reached this latter distance, the rod has moved away from S an additional distance equal to $\frac{v_0}{C}\left(\frac{v_0}{C}\ell\right)$ and so on.

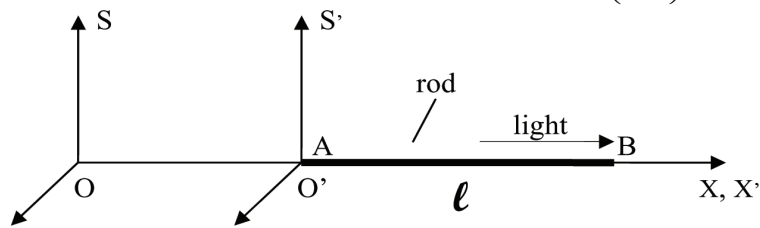

Figure 5. The co-ordinate system S is attached to the Cosmic substratum (aether frame). S' moves away from S with rectilinear motion along the common x axis. At the initial instant O and O' are coincident. At this instant a light ray emanates from the common origin and travels toward point B.

So that, the total distance covered by the ray in S when it reaches the end of the rod is:

$$\ell\left(1+\frac{v_0}{C}+\frac{v_0^2}{C^2}+...+\frac{v_0^n}{C^n}+...\right)=\ell\sum_{i=1}^{\infty}1+\frac{v_0^i}{C^i}$$

The sum of the series is: $\ell\dfrac{C}{C-v_0}$

Let us name $L$ the length of the rod when it is at rest in the co-ordinate system S. Due to Lorentz contraction it is reduced in S' to:
$$\ell = L\sqrt{1-v_0^2/C^2}$$
The distance covered by the ray in S will therefore be:
$$x = L\frac{C}{C-v_0}\sqrt{1-v_0^2/C^2} = \frac{L+\dfrac{v_0 L}{C}}{\sqrt{1-v_0^2/C^2}}$$

From this expression, we easily obtain the transit time of the signal according to observer S: 
$$t = \frac{x}{C} = \frac{L/C + v_0 L/C^2}{\sqrt{1-v_0^2/C^2}} \qquad (3)$$

In the ideal case where the measurements are carried out perfectly by observer S', this observer finds also for $t'$
$$t' = \frac{L\sqrt{1-v_0^2/C^2}}{C-v_0} = \frac{L/C + v_0 L/C^2}{\sqrt{1-v_0^2/C^2}} \qquad (4)$$

So that $t = t'$. But, as we have seen before, when observer S' measures the speed of light, he makes a systematic error and finds $C$ (consult formula (2) and the important remark following the formula). This is also the case for the length: since the rod is measured with a contracted meter stick, its length in S' is erroneously found equal to $L$. Therefore, for observer S' the *apparent* time needed by the ray to reach the end of the rod is

$$t'_{app} = \frac{L}{C} \qquad (5)$$

Since the length $L$ is arbitrary: $L = x'_{app}$. Then comparing $t$ and $t'_{app}$ gives:

$$t = \frac{t'_{app} + v_0 \dfrac{x'_{app}}{C^2}}{\sqrt{1-v_0^2/C^2}} \qquad (6)$$

We see that, contrary to the assertions of special relativity, $t'_{app}$ is a fictitious apparent time. Only $t$ is the real time. Nevertheless, $t'_{app}$ is the time measured by observer S'. From expression (6) we easily obtain:

$$x = \frac{x'_{app} + v_0 t'_{app}}{\sqrt{1 - v_0^2/C^2}} \tag{7}$$

and the reciprocal transformations:

$$x'_{app} = \frac{x - v_0 t}{\sqrt{1 - v_0^2/C^2}} \tag{8} \quad \text{and} \quad t'_{app} = \frac{t - v_0 x/C^2}{\sqrt{1 - v_0^2/C^2}} \tag{9}$$

These reciprocal transformations can be qualified as Lorentz-Poincaré transformations since $x$ and $t$ are the real coordinates relative to the fundamental frame. Yet the measured time and distance in frame S' have a different meaning than in conventional relativity, they are the (*apparent*) co-ordinates and not the real co-ordinates; the real co-ordinates when the light ray reaches point B being $x' = x - v_0 t = \ell$ and $t' = t$.

*Important remarks*

- We see that the formulas (3) and (4) are correct since $t = t'$. But insofar as $t'_{app} = \frac{L}{C}$ and $x'_{app} = L$ are considered the real coordinates, we are misled.

- If the clocks had been perfectly synchronized, then, as a result of the slowing down of moving clocks, we would have obtained (from formula (6)):

$$t'_{app} = t\sqrt{1 - v_0^2/C^2} \tag{10}$$

(and not $t'_{app} = t$ which does not imply the slowing down of the clocks in S')

Formula (10) is the expression used by Tangherlini [12] and Mansouri and Sexl [13].

The above transformations (6) to (9), look compatible with the relativity principle, and imply reciprocity of observations (*apparent*). But according to aether theory, there is no real reciprocity. For example, the rods at rest in the fundamental frame do not contract. The apparent reciprocity results from the impossibility of synchronizing exactly the clocks in S' *by means of the usual methods*. This apparent reciprocity has been demonstrated by Prokhovnik [8]. It was used as an argument to prove that the aether of Lorentz is compatible with the principle of relativity. Nevertheless, the argument is not sufficient as the examples given in this text and in ref [1a, 1c, 1d] demonstrate.

**V.1.2** Let us now study a different case: as figure 6 shows, it deals with two 'inertial systems' $S_1$ and $S_2$, receding from the fundamental system $S_0$ along the common *x*-axis. (This case is important because it is the case generally observed in practice).

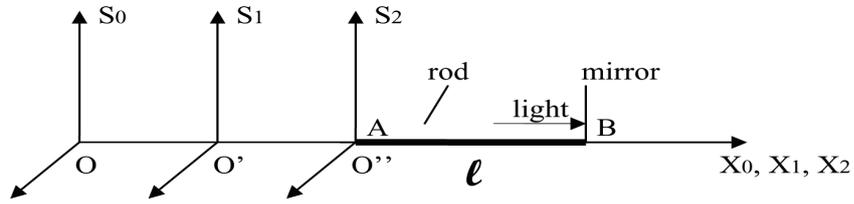

Figure 6. Both co-ordinate systems $S_1$ and $S_2$ are in motion with respect to the aether frame. At the initial instant, the origin of the three co-ordinate systems O, O' and O'' are coincident. When the signal coming from the common origin reaches point B, it meets a mirror firmly attached to frame $S_1$.

The relative speeds between the co-ordinate systems are $v_{01}$, $v_{02}$ and $v_{12}$. A long rod AB firmly fixed to the $x_2$-axis of $S_2$ which measured $\ell_0$ when it was at rest in $S_0$, measures $\ell = \ell_0\sqrt{1 - v_{02}^2/C^2}$ in $S_2$. At the initial instant $t_0 = $ zero, the origins O, O' and O'' of the three reference systems and the origin A of the rod are coincident. At this very instant, a light ray emanates from this origin, and travels along the common *x*-axis towards point B. When the signal reaches point B, it meets a mirror *firmly attached to the system* $S_1$. The distance between O' and the mirror is therefore constant. After reflection, the signal comes back to O'. As we have seen (formulas (3) and (4)), for observer $S_0$ the time needed by the signal to reach point B is

$$t_0 = \frac{\ell_0\sqrt{1 - v_{02}^2/C^2}}{C - v_{02}} \qquad (11)$$

According to Lorentz, this time is the real transit time of the signal. Now, let us determine the *apparent* transit time measured by observer $S_1$.

Assuming that the speed of light is $C - v_{01}$ in $S_1$ and $C - v_{02}$ in $S_2$, the real transit time $t_1$ from O' to B, can be easily obtained. Let us first determine the space co-ordinate. We can see that, when the ray has covered in $S_1$ a distance equal to $\ell$, $S_2$ has moved away from $S_1$ a distance equal to: $\dfrac{v_{12}}{C - v_{01}} \ell$

When the ray has covered this distance in its turn, $S_2$ has moved away from $S_1$ an additional distance equal to: $\dfrac{v_{12}}{C - v_{01}} \left( \dfrac{v_{12}}{C - v_{01}} \right) \ell = \dfrac{v_{12}^2}{(C - v_{01})^2} \ell$ and so on. Therefore the distance $x_{1r}$ covered by the ray in $S_1$ when it reaches the end of the rod is

$$\ell_0 \sqrt{1 - v_{02}^2 / C^2} \left( 1 + \dfrac{v_{12}}{C - v_{01}} + \dfrac{v_{12}^2}{(C - v_{01})^2} + \ldots + \dfrac{v_{12}^n}{(C - v_{01})^n} + \ldots + \ldots \right)$$

$$\ell_0 \sqrt{1 - v_{02}^2 / C^2} \, \dfrac{C - v_{01}}{C - v_{01} - v_{12}} = \ell_0 \sqrt{1 - v_{02}^2 / C^2} \, \dfrac{C - v_{01}}{C - v_{02}} \tag{12}$$

(Note that, if we assume the Lorentz postulates, real speeds obey the Galilean addition of velocities law. The relativistic law applies when the usual measurement procedures, which alter the measurement, are used - see later).

Now, the measurement of the distance O'B by observer $S_1$ is carried out with a contracted meter stick, so that he will find (in place of expression (12)):

$$x_{1app} = \ell_0 \dfrac{\sqrt{1 - v_{02}^2 / C^2}}{\sqrt{1 - v_{01}^2 / C^2}} \dfrac{C - v_{01}}{C - v_{02}}$$

The real transit time of the signal (for $S_1$) from O' to B is (from (12)):

$$t_1 = \ell_0 \dfrac{\sqrt{1 - v_{02}^2 / C^2}}{C - v_{02}} = \dfrac{\ell_0 / C + v_{02} \ell_0 / C^2}{\sqrt{1 - v_{02}^2 / C^2}}$$

Contrary to what special relativity asserts, it is the same as the time $t_0$ measured by observer $S_0$ (see formula (11)). According to Lorentz, it is the universal transit time (which is the same in all 'inertial frames'). Now, as we have seen, there are great difficulties in measuring $t_1$. The time generally measured in place of $t_1$ (with the retarded clocks attached to frame $S_1$) is the *apparent* average transit time: $\tau_{1app} = \dfrac{t_1 + \bar{t_1}}{2}\sqrt{1 - v_{01}^2/C^2}$ (see important remark at the end of section IV, and [1]), where $\bar{t_1}$ is the time needed by the signal to return from the mirror to O'.

The distance covered by the ray from the mirror to O' is the same as from O' to B, but the speed of light in the reverse direction (with respect to $S_1$) is $C + v_{01}$.

So that $\bar{t_1} = \ell_0 \sqrt{1 - v_{02}^2/C^2}\, \dfrac{C - v_{01}}{C - v_{02}} \times \dfrac{1}{C + v_{01}}$

Finally: $\tau_{1app} = \dfrac{1}{2}\ell_0 \sqrt{1 - v_{02}^2/C^2}\, \dfrac{C - v_{01}}{C - v_{02}} \times \dfrac{2C}{C^2 - v_{01}^2}\sqrt{1 - v_{01}^2/C^2}$

$$= \dfrac{\ell_0}{C}\, \dfrac{\sqrt{1 - v_{02}^2/C^2}}{\sqrt{1 - v_{01}^2/C^2}}\, \dfrac{C - v_{01}}{C - v_{02}} \qquad (13)$$

$\tau_{1app} = \tau_{2app}\, \dfrac{\sqrt{1 - v_{02}^2/C^2}}{\sqrt{1 - v_{01}^2/C^2}}\, \dfrac{C - v_{01}}{C - v_{02}}$ (From formula (5))

$\tau_{1app}$ is the *apparent* average transit time of light measured with clocks attached to the system $S_1$. If we suppose that $\tau_{1app}$ and $x_{1app}$ are the real co-ordinates of the signal when it reaches point B, then the speed of light in $S_1$ is (erroneously) found to be of magnitude $C$.

We note that for $v_{01} = 0$ we obtain:

$$x_0 = x_{2app}\sqrt{1 - v_{02}^2/C^2}\, \dfrac{C}{C - v_{02}} = \dfrac{\ell_0 + v_{02}\ell_0/C}{\sqrt{1 - v_{02}^2/C^2}}$$

$$\tau_0 = \tau_{2app}\sqrt{1 - v_{02}^2/C^2}\, \dfrac{C}{C - v_{02}} = \dfrac{\ell_0\sqrt{1 - v_{02}^2/C^2}}{C - v_{02}} = \dfrac{\dfrac{\ell_0}{C} + v_{02}\dfrac{\ell_0}{C^2}}{\sqrt{1 - v_{02}^2/C^2}}$$

And therefore:

$x_{2app} = \dfrac{x_0 - v_{02}\tau_0}{\sqrt{1 - v_{02}^2/C^2}}$

$$\tau_{2app} = \frac{\tau_0 - v_{02}\dfrac{x_0}{C^2}}{\sqrt{1 - v_{02}^2/C^2}}$$

This is in conformity with our expectations: when the speed of the system $S_1$ is zero, $S_1$ is at rest in the cosmic substratum and the Lorentz-Poincaré transformations apply. (Only the interpretation of the co-ordinates in $S_2$ is different from conventional relativity. For aether theory $x_{2app}$ and $\tau_{2app}$ are not the real co-ordinates, they are the measured (*apparent*) distance and time).

For $v_{01} = v_{02}$ we have $\tau_{1app} = \dfrac{\ell_0}{C}$,

in this case $S_1$ and $S_2$ are not different and $\tau_{1app} = \tau_{2app}$.

**V.2 Transformations of space and time for bodies moving uniformly at speeds lower than the speed of light.**

Consider the three 'inertial systems' previously mentioned and let a vehicle coming from the $-x$ direction, pass at the initial instant ($t_0 = 0$) by the common origin (O, O', O") (Figure 7). The vehicle travels uniformly along a rigid path AB firmly fixed to the $x_2$-axis of the system $S_2$. The origin A of the path permanently coincides with point O".

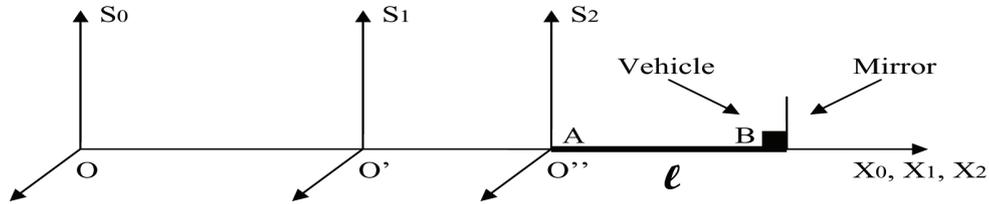

Figure 7. When the vehicle reaches point B, it meets a clock equipped with a mirror firmly attached to the co-ordinate system $S_1$.

When the vehicle reaches point B, it meets a clock $C\ell$ equipped with a mirror *firmly attached to* $S_1$. We propose to compare the relative distances covered by the vehicle in $S_1$ and $S_2$ and the transit times measured by the observers standing in these co-ordinate systems by means of the usual methods of measurement.

The real distance $X_{1r}$ covered by the vehicle in $S_1$ can be easily obtained: indeed, the ratio of the distances covered in $S_1$ and $S_2$ is equal to the ratio of the speeds with respect to these two co-ordinate systems. That is:

$$\frac{X_{1r}}{\ell_0 \sqrt{1 - v_{02}^2 / C^2}} = \frac{V - v_{01}}{V - v_{02}}$$

where $V$ represents the real speed of the vehicle with respect to $S_0$. $v_{01}$ and $v_{02}$ are the real speeds of $S_1$ and $S_2$ with respect to $S_0$. (Notice that real speeds are simply additive. As we shall see, only the apparent speeds obey a law of composition different from the Galilean law).

So: $X_{1r} = \ell_0 \sqrt{1 - v_{02}^2 / C^2} \dfrac{V - v_{01}}{V - v_{02}}$ (with $V > v_{02}$) \hfill (14)

Since the distance separating point O' and the mirror is measured with a contracted meter stick, the apparent distance $X_{1app}$ found by observer $S_1$ will be:

$$X_{1app} = \frac{\ell_0 \sqrt{1 - v_{02}^2 / C^2}}{\sqrt{1 - v_{01}^2 / C^2}} \frac{V - v_{01}}{V - v_{02}} \tag{15}$$

Now, in order to measure in $S_1$ the time needed by the vehicle to reach point B, we must beforehand synchronize the clocks O' and $C\ell$.

As we have seen in the previous chapters, the method of Poincaré-Einstein considers that the clock display $\dfrac{t_1 + \bar{t}_1}{2} \sqrt{1 - v_{01}^2 / C^2}$ is the one-way transit time of light. In reality, it is the 'apparent' average transit time of light $\tau_{1app}$. The real transit time of light from O' to the clock $C\ell$ is in fact: $t_1 = \dfrac{X_{1r}}{C - v_{01}}$ \hfill (16)

and from the clock $C\ell$ to O': $\bar{t}_1 = \dfrac{X_{1r}}{C + v_{01}}$ \hfill (17)

Taking account of clock retardation in $S_1$, the synchronism discrepancy $\Delta$ between the clocks O' and $C\ell$ is therefore given by: (see ref [1])

$$\Delta = t_1 \sqrt{1 - v_{01}^2 / C^2} - \frac{(t_1 + \bar{t}_1)}{2} \sqrt{1 - v_{01}^2 / C^2} = \frac{(t_1 - \bar{t}_1)}{2} \sqrt{1 - v_{01}^2 / C^2}$$

From (14), (16), and (17) we obtain:

$$\Delta = \frac{v_{01} \ell_0}{C^2} \frac{\sqrt{1 - v_{02}^2 / C^2}}{\sqrt{1 - v_{01}^2 / C^2}} \frac{V - v_{01}}{V - v_{02}}$$

Now, the real time needed by the vehicle to cover the distance $X_{1r}$ is:

$$T_{1r} = \frac{X_{1r}}{V - v_{01}} = \frac{\ell_0 \sqrt{1 - v_{02}^2/C^2}}{V - v_{02}} \qquad \text{(from (14))}$$

This time is the universal time that clocks would display if they were at rest in the aether frame (in which there is no speed of light anisotropy and no clock retardation).

But in $S_1$ we must take account of the synchronism discrepancy effect and of clock retardation, so that the experimental *apparent* time obtained when we use the method of Poincaré-Einstein is:

$$\begin{aligned} T_{1app} &= T_{1r}\sqrt{1 - v_{01}^2/C^2} - \Delta \\ &= \ell_0 \frac{\sqrt{1 - v_{02}^2/C^2}}{\sqrt{1 - v_{01}^2/C^2}} \frac{(1 - v_{01}V/C^2)}{V - v_{02}} \end{aligned} \qquad (18)$$

From expressions (15) and (18) we obtain:

$$V_{1app} = \frac{X_{1app}}{T_{1app}} = \frac{V - v_{01}}{1 - v_{01}V/C^2} \qquad (19)$$

This expression takes the same form as the composition of velocities law of special relativity but, obviously, it has not the same meaning.

Expressions (15) and (18) can be expressed as functions of $T_{2app}$ and $X_{2app}$. We note that the length of the path is arbitrary, and since it is measured in $S_2$ with a contracted standard, we have $X_{2app} = \ell_0$

we also note that $X_{2app} = \frac{V - v_{02}}{1 - v_{02}V/C^2} T_{2app}$

replacing $\ell_0$ with this expression in (18) we obtain

$$T_{1app} = T_{2app} \frac{\sqrt{1 - v_{02}^2/C^2}}{\sqrt{1 - v_{01}^2/C^2}} \frac{(1 - v_{01}V/C^2)}{(1 - v_{02}V/C^2)} \qquad (20)$$

and replacing $\ell_0$ with $X_{2app}$ in (15) gives

$$X_{1app} = X_{2app} \frac{\sqrt{1 - v_{02}^2/C^2}}{\sqrt{1 - v_{01}^2/C^2}} \frac{V - v_{01}}{V - v_{02}} \qquad (21)$$

We can now see that, contrary to Einstein's special relativity, $v_{01}$ and $v_{02}$, which are the velocities of $S_1$ and $S_2$ with respect to the aether frame, are systematically omnipresent in the equations.

Expressions (20) and (21) are the extended space time transformations, applicable to any pair of 'inertial bodies' receding uniformly with respect to one another along the direction of motion of the solar system. It should be pointed out that during a brief interval of time, the motion of the Earth with respect to the Cosmic Substratum can be considered rectilinear and uniform. If this was not the case, the bodies placed on its surface would be subjected to perceptible accelerations.

With respect to the absolute motion of the sun, the orbital and the rotational motions of the Earth are slow. So, as a first approximation and during a short time, the motion of the Earth can be identified with the motion of the solar system. This means that $S_1$ can be identified with the Earth frame, and $S_2$ with a vehicle moving on its surface in the direction of motion of the Solar system (a ship for example). Accordingly, $v_{01}$ is the real speed of the Earth with respect to the Cosmic Substratum $S_0$, $v_{02}$ is the real speed of the ship with respect to $S_0$.

$V$ is the real speed, with respect to $S_0$, of a body present on the surface of the ship and moving in the same direction, and $V_{1app}$ is the apparent speed of the body with respect to the Earth frame.

(Formula (21) is identical to the expression of the space transformation which was given in a previous publication [36]. Formula (20) represents a completely satisfactory expression of the time transformation: it applies to all values of $V$. It replaces the expression given in [36] which was limited to high values of $V$).

Note

It is interesting to also express $V_{1app}$ as a function of $V_{2app}$

From $V_{2app} = \dfrac{V - v_{02}}{1 - v_{02}V/C^2}$ we obtain

$$V = \dfrac{V_{2app} + v_{02}}{1 + v_{02}V_{2app}/C^2}$$

Replacing this expression of $V$ in (19) we obtain:

$$V_{1app} = \dfrac{\dfrac{V_{2app} + v_{02}}{1 + v_{02}V_{2app}/C^2} - v_{01}}{1 - \dfrac{v_{01}}{C^2}\left(\dfrac{V_{2app} + v_{02}}{1 + v_{02}V_{2app}/C^2}\right)}$$

We remark that, in conformity with the *apparent* (experimental) speed of light invariance, $V_{1app} = C \Rightarrow V_{2app} = C$ and conversely.

**Coherence of the derivation**

The coherence of the derivation can be checked. To this end, we need to demonstrate its agreement with known experimental data. The equations must reduce to the Lorentz-Poincaré transformations when one of the co-ordinate systems they connect is at rest in the Cosmic Substratum (see below), and they must explain why the experimental measurement of the speed of light by the usual methods, always gives $C$.

We note that when $v_{01} = v_{02}$, the systems $S_1$ and $S_2$ which were coincident at the initial instant, always remain coincident. In this case, as expected, $\Delta$ is reduced to $v_{01}\ell_0/C^2$ which is the synchronism discrepancy effect defined by Prokhovnik [8]. (More exactly Prokhovnik takes $\delta = \frac{v_{01}\ell_0}{C^2}\gamma$ as a definition of the concept. But, as demonstrated in [1] and [31], contrary to the opinion of this author, this implies that the measurement does not take account of the slowing down of clocks caused by the movement. See also the chapter VII)

For $V = C$, the apparent time and space co-ordinates reduce to:

$$T_{1app} = \frac{\ell_0}{C} \frac{\sqrt{1-v_{02}^2/C^2}}{\sqrt{1-v_{01}^2/C^2}} \frac{C-v_{01}}{C-v_{02}}$$

and $X_{1app} = \ell_0 \frac{\sqrt{1-v_{02}^2/C^2}}{\sqrt{1-v_{01}^2/C^2}} \frac{C-v_{01}}{C-v_{02}}$

We remark that the *apparent* (measured) speed of light $V_{1app}$ in $S_1$ is equal to $C$ in conformity with the experiment.

Now, when $v_{01} = 0$, $S_1$ is at rest in the Cosmic Substratum and then

$$X = \ell_0 \frac{\sqrt{1-v_{02}^2/C^2}}{V-v_{02}} V$$

and $T = \ell_0 \frac{\sqrt{1-v_{02}^2/C^2}}{V-v_{02}}$

$$T = \frac{\ell_0\sqrt{1-v_{02}^2/C^2}}{C-v_{02}} \frac{C-v_{02}}{V-v_{02}}$$

$$= \frac{\ell_0/C + v_{02}\ell_0/C^2}{\sqrt{1-v_{02}^2/C^2}} \frac{C-v_{02}}{V-v_{02}}$$

After multiplication of the two fractions we obtain:

$$T = \frac{\ell_0 - v_{02}^2 \ell_0/C^2}{\sqrt{1-v_{02}^2/C^2}\ (V-v_{02})}$$

$$= \frac{\ell_0(1-v_{02}V/C^2) + (V-v_{02})v_{02}\ell_0/C^2}{\sqrt{1-v_{02}^2/C^2}\ (V-v_{02})}$$

Taking account of the fact that

$$V_{2app} = \frac{V-v_{02}}{1-v_{02}V/C^2} \quad \text{and} \quad X_{2app} = \ell_0$$

we obtain $T = \dfrac{T_{2app} + v_{02}X_{2app}/C^2}{\sqrt{1-v_{02}^2/C^2}}$

and $X = \dfrac{T_{2app} + v_{02}X_{2app}/C^2}{\sqrt{1-v_{02}^2/C^2}} \times V$ \hfill (22)

Now, from the expression $V = \dfrac{V_{2app} + v_{02}}{1 + v_{02}V_{2app}/C^2}$

we easily find from (22) that: $X = \dfrac{X_{2app} + v_{02}T_{2app}}{\sqrt{1-v_{02}^2/C^2}}$

From these expressions we obtain:

$$T_{2app} = \frac{T - v_{02}X/C^2}{\sqrt{1-v_{02}^2/C^2}}$$

$$X_{2app} = \frac{X - v_{02}T}{\sqrt{1-v_{02}^2/C^2}}$$

Therefore, as expected, we obtain the Lorentz-Poincaré transformations when one of the co-ordinate systems is at rest in the Cosmic Substratum, (the only difference with conventional relativity rests on the interpretation of the time and the distance in the co-ordinate system $S_2$.)

 *- Important remarks*

$\ell_0$ is not the real co-ordinate of point B relative to $S_2$ along the $x_2$-axis, the real co-ordinate is $\ell$. The ignorance of this fact is a source of much confusion.

It should also be pointed out that, contrary to what is often believed, $X_{1app}$, $T_{1app}$ and $V_{1app}$ are all apparent (fictitious) co-ordinates.

**VI. Variation of mass with speed in relativity and in the fundamental aether theory.**

In the present chapter, we propose to test the coherence of special relativity in the field of dynamics, such as the laws are assumed, without calling into question *a priori* the compatibility of the relativity principle with mass-energy conservation, a question which will be tackled in the following chapter.

In relativity, since no absolute frame exists, the mass of a body attached to any given 'inertial frame', viewed by an observer at rest in this frame, is always regarded as identical. This mass is defined as the proper mass or the rest mass of the body. If the body moves with respect to a frame S with velocity *v*, its mass with respect to S is assumed to be:

$$m = \frac{m_0}{\sqrt{1 - v^2/C^2}}$$

whatever the reference frame S may be.

The point of view of the fundamental aether theory is completely different. Indeed, consider a body having the mass $m_0$ in the fundamental frame $S_0$. Since the body needs to acquire kinetic energy $E_C$ in order to move from frame $S_0$ to any other 'inertial frame' $S_1$, the rest mass of the body in this frame will be $m_0 + \frac{E_C}{C^2}$. This means that a hierarchy of rest masses exists, each a function of the absolute speed of the body.

(Note that it is necessary to distinguish the real mass from the measured mass, which can be incorrectly determined. Indeed, if one measures the mass $m_0$ of a body in the fundamental frame by comparison with a standard $\mu_0$, if $m_0$ and $\mu_0$ are transported in another 'inertial frame', they modify in the same ratio. As a result, the mass $m_0$ does not appear to have changed, which is inexact).

As we have seen in the previous chapters and in [1, 35, 39], contrary to what is often claimed, the existence of a fundamental frame is not compatible with the exact application of the relativity principle.

We shall also verify this in the following example. Consider three 'inertial frames' $S_0$, $S_1$ and $S_2$, and let three bodies of masses $m_0$, $m_1$ and $m_2$ be respectively at rest in these three frames. The said masses were initially identical in reference frame $S_0$ and equal to $m_0$, before being transported in their respective reference frame. We propose to examine the effect of motion on these masses (see figure 8).

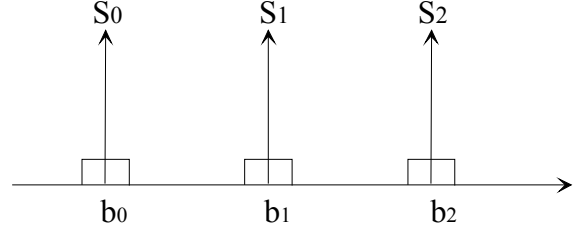

Figure8. The masses of the three bodies were identical in frame $S_0$ before being transported in their respective reference frame.

**VI.1 Point of view of the conventional theory of relativity.**
Measured by an observer at rest with respect to one of the bodies, its mass remains, in all cases, equal to $m_0$. Therefore, for observer $S_1$, we have

$$m_2^1 = \frac{m_0}{\sqrt{1 - v_{12}^2/C^2}} \quad (23)$$

Where $m_2^1$ refers to the relativistic mass of body $b_2$ measured by observer $S_1$ and $v_{12}$ refers to the relative speed between the reference frames $S_1$ and $S_2$. If one supposes that $v_{12} \ll C$, expression (23) can be written to first order as follows

$$m_2^1 \cong m_0\left(1 + \frac{1}{2}v_{12}^2/C^2\right) \quad (24)$$

So that, viewed by observer $S_1$, the energy of body $b_2$ is

$$m_2^1 C^2 \cong m_0 C^2 + \frac{1}{2}m_0 v_{12}^2$$

(This corresponds to the sum of the rest energy and the kinetic energy needed by $b_2$ to move from $S_1$ to $S_2$).

For an observer attached to reference frame $S_0$, the energy of $b_2$ is different. Assuming that $m_2^0$ refers to the mass of body $b_2$ measured by this observer we have, (for $v_{02} \ll C$):

$$m_2^0 C^2 \cong m_0 C^2 + \frac{1}{2}m_0 v_{02}^2$$

and the energy of body $b_1$ is assumed to be

$$m_1^0 C^2 \cong m_0 C^2 + \frac{1}{2} m_0 v_{01}^2$$

So that, for observer $S_0$ the kinetic energy needed by the body $b_2$ to move from $S_1$ to $S_2$ is

$$(m_2^0 - m_1^0) C^2 \cong \frac{1}{2} m_0 (v_{02}^2 - v_{01}^2)$$

This result is different from the measurement made by observer $S_1$, $\frac{1}{2} m_0 v_{12}^2$, although, obviously, it should be the same.

**VI.2 Point of view of the fundamental aether theory**

In our book "Relativité et substratum cosmique [37]", the results that will follow were seen as a stumbling block for the fundamental aether theory, because they lead to an expression for kinetic energy different from the usual expression. Nevertheless, objections to the application of the relativity principle and present day arguments in favour of the aether and of the anisotropy of the one-way speed of light compel us to reassess our earlier point of view.

Let us now go back to the figure with the three bodies, and suppose that $S_0$ is the fundamental inertial frame, and $S_1$ and $S_2$ two 'inertial frames' aligned with $S_0$. $m_0$ refers to the mass of the bodies in frame $S_0$.

According to the fundamental aether theory, $m_2^0$ and $m_2^1$ have no meaning. A body at rest in a given 'inertial frame' has only one real mass. The mass of the body $b_2$ is:

$$m_2 = \frac{m_0}{\sqrt{1 - v_{02}^2 / C^2}} \tag{25}$$

and the mass of $b_1$:

$$m_1 = \frac{m_0}{\sqrt{1 - v_{01}^2 / C^2}} \tag{26}$$

Conversely, as we shall see, the rest mass of a body will not be $m_0$ in the different 'inertial frames', from (25) and (26) we obtain

$$m_2 = m_1 \frac{\sqrt{1 - v_{01}^2 / C^2}}{\sqrt{1 - v_{02}^2 / C^2}} \tag{27}$$

If one supposes that $v_{02} \ll C$, $m_2$ reduces to

$$m_2 \cong m_1 + \frac{m_1}{2C^2}\left(v_{02}^{\ 2} - v_{01}^{\ 2}\right) \tag{28}$$

$$\cong m_1 + \frac{m_1}{2C^2}\left(v_{12}^{\ 2} + 2v_{01}v_{12}\right)$$

In this case $m_1$ is hardly different from $m_0$.

The kinetic energy acquired by a body which moves from frame $S_1$ to $S_2$ is $m_2 C^2 - m_1 C^2$. It reduces to $1/2 m_0 (v_{02}^2 - v_{01}^2)$ when $v_{02} \ll C$. It is different from $1/2 v_{12}^2$.

We may notice that expression (27), which connects any pair of 'inertial frames', does not assume a mathematical form identical to (25) and (26), in contradiction with the application of the relativity principle.

We also note that, when $v_{12} \to 0$ or in other words when $v_{02} \to v_{01}$, the terms depending on $v_{01}$ and $v_{02}$ in expression (28) cancel. Thus, $m_1$ represents the rest mass assumed by the aforementioned bodies when they stand in reference frame $S_1$. This is different from special relativity for which the rest mass is $m_0$ in any 'inertial frame'.

Nevertheless, we must distinguish the absolute rest mass $m_0$ from the other rest masses standing in 'inertial frames' that are in motion with respect to the aether frame.

Note however that when $v_{12} \gg v_{01}$, and $v_{01} \ll C$ expression (27) reduces to

$$m_2 \cong \frac{m_1}{\sqrt{1 - v_{02}^{\ 2}/C^2}} \cong \frac{m_1}{\sqrt{1 - v_{12}^{\ 2}/C^2}}$$

and since $m_1 \cong m_0$, we obtain

$$m_2 \cong \frac{m_0}{\sqrt{1 - v_{12}^{\ 2}/C^2}}$$

It is the case of particles moving at very high speed $(v \cong c)$ (while the Earth moves with respect to the aether frame at *relatively* low speed ($\cong 350$ km/sec)). In such cases, as a first approximation, the fact that the Earth has absolute motion, hardly affects the results of the calculations. So relativity and aether theory lead to practically equivalent results.

### VI.3 The question of reciprocity.

This question makes a great difference between relativity and the fundamental aether theory. According to relativity, when a body is transported from one 'inertial system' $S_0$, to another $S_1$, viewed from $S_0$ its mass is supposed to be

$$m_1 = \frac{m_0}{\sqrt{1 - v_{01}^2/C^2}}$$

But, conversely, if the body comes back to $S_0$, viewed from $S_1$ its mass will also be considered equal to $m_1$

For the treatment in the fundamental aether theory, let us assume that $S_0$ is the fundamental frame. If the mass is at rest in frame $S_1$, we also have

$$m_1 = \frac{m_0}{\sqrt{1 - v_{01}^2/C^2}}$$

$m_1$ is $> m_0$, indeed we have been compelled to supply energy to the body in order to move it from $S_0$ to $S_1$, but if the body returns to $S_0$, the energy is restored. All observers (including the observer attached to frame $S_1$) will conclude that the real mass of the body in frame $S_0$ is equal to $m_0$.

This result is in total contradiction with relativity, but it is the only result which is in accordance with mass-energy conservation.

<u>Important remark.</u>

In the fundamental aether theory, we must distinguish the total available energy of a body (which is equal to the sum of the rest energy $m_0 C^2$ and the kinetic energy with respect to the fundamental frame), from the available energy of the body with respect to any other 'inertial frame', which is smaller than the previous energy, and takes another mathematical form.

In the example previously quoted, the total available energy of body $b_2$ is

$$m_2 C^2 = m_0 C^2 \left(1 + \frac{1}{2} v_{02}^2/C^2\right) + \text{small terms of higher order}$$

(This notion has no equivalence in conventional relativity for which the energy of a body is completely relative and depends on its speed with respect to another body).

**VI.4 Possible measurement of the absolute speed of an 'inertial system'**

Assuming that $v_{02} \ll C$, the kinetic energy needed to move a body from $S_1$ to $S_2$ reduces to:

$$\Delta E_C \cong \frac{1}{2} m_0 \left(v_{12}^2 + 2 v_{01} v_{12}\right)$$

Knowing $\Delta E_C$ and $v_{12}$, it is theoretically possible to measure the absolute speed $v_{01}$ of the 'inertial system' $S_1$, that is:

$$v_{01} \cong \frac{\Delta E_C - \frac{1}{2}m_0 v_{12}^2}{m_0 v_{12}}$$

This result is also in contradiction with the relativity principle.

## VII. Inferences from the aether theory

Some of the topics treated here had been partly tackled in previous publications. Yet the reader will find additional explanations in response to the questions asked to the author.

### The relativity principle

The idea of relativity has played an important role in the development of contemporary physics. It has been used successfully by Galileo to refute some dogmas of Aristotelian physics. The arguments of Galileo appeared so convincing that the idea has acquired, in its turn, the status of an irrefutable dogma. In the light of the new data presented in this manuscript, it is appropriate to go through its conceptual content, once more.

Aristotelian physics considered motion and rest as absolute and completely distinct, the Earth, in a state of absolute rest, occupying the centre of the Universe. Galileo departed completely from this idea. For him, uniform motion presented no absolute character. Any object at rest with respect to a given reference system is, at the same time, in motion with respect to another reference system; rest and motion are not fundamentally different, they have only relative character.

As an example of his relativity principle, Galileo cited the case of a stone released from the top of a ship's mast, as the ship sails uniformly in a straight line. If, according to Galileo, motion had an absolute character, the stone would fall at a distance from the foot of the mast. But the stone falls at the foot of the mast, a result he interpreted as due to the fact that the ship is at rest in its 'inertial system'.

Although the argument of Galileo demonstrated the acuteness of his critical judgment, we must be aware that the knowledge, in his time, could not allow him to study the problem from all its sides. His argument is in fact questionable. Indeed, the stone has momentum $\overrightarrow{mv}$ parallel to the direction of motion of the ship, and the fact that it is released does not eliminate this momentum which constrains it to continue its horizontal motion along with the ship, while its vertical motion is determined by the law of gravitation.

Therefore, even if motion and rest present an absolute character, the stone will still fall at the foot of the mast.

(Note that our reasoning is entirely exact only in the ideal case where the motion of the ship is rectilinear and uniform, and where air resistance is negligible. To be rigorous we should also take account of the resistance opposed by the substratum. But in the present example this resistance would be very small).

The fact remains that it is to Galileo's credit that the meaning of motion is clearly understood. Contrary to Aristotle's belief, Galileo demonstrated that if motion needs a motor to be produced, it does not need a motor to be maintained.

-Galilean relativity principle was limited to the laws of mechanics and to bodies moving at low speeds. Poincaré's purpose was to extend it to all the laws of physics and in particular to Maxwell's electromagnetism. Indeed, these laws seemed to be an exception to the rule. In order to bring them back into line, Poincaré had to resort to a set of equations he baptised Lorentz transformations, which should assume a group structure.

Poincaré expressed his principle in the following terms: "It seems that the impossibility of observing the absolute motion of the Earth is a general law of nature. We are naturally inclined to admit this law that we shall call relativity postulate, and to admit it without restriction [38]".

This sentence does not imply dismissal of absolute motion but rather the impossibility of observing it. However, on another occasion, Poincaré declared: "there is no absolute space; all the motions we can observe are relative motions [38]".

Nevertheless, Poincaré did not reject the aether. His acceptance of this medium is expressed in the following terms:

"Does an aether really exist? The reason why we believe in an aether is simple: if light comes from a distant star and takes many years to reach us, it is, during its travel, no longer on the star, but not yet near the Earth. Nevertheless it must be somewhere and supported by a material medium" (La science et l'hypothèse, chapter 10, p 180 of the French edition "Les theories de la physique moderne".)

During a lecture given in Lille (France) in 1909 [38] Poincaré declared:

"Let us remark that an isolated electron moving through the aether generates an electric current, that is to say an electromagnetic field. This field corresponds to a certain quantity of energy localized in the aether rather than in the electron.

Poincaré acknowledged his debt to Lorentz in the following terms:

"The results I have obtained agree with those of Mr. Lorentz in all important points. I was led to modify and complete them in a few points of detail."

Therefore Poincaré tried to reconcile the Lorentz aether with relativity.

Poincaré's point of view cannot be maintained. We have demonstrated this in the example of two vehicles moving in opposite directions towards two symmetrical targets (see end of chapter II of this text and ref [23]). In addition, since Poincaré assumes length contraction, a rod in motion with respect to the aether frame will shrink when its orientation will be changed (from the perpendicular to the parallel direction of motion). Of course we will not be able to measure this contraction with a meter stick, since the stick also shrinks in the same ratio. But, at high speed, the process will be observable and will inform us about the absolute motion of the rod. (Note also that, with four meter sticks in motion at high speed, two of them being parallel to the direction of motion and two others perpendicular, we will construct a rectangle instead of a square).

So, contrary to Poincaré's view, the aether of Lorentz is not compatible with the exact applicability of the relativity principle in the physical world.

Here, it is important to specify that we don't contest the relativity principle as an abstract concept. Indeed, if moving bodies were not under the influence of any physical forces, the relativity principle would strictly apply. But insofar as there is an aether drift whose magnitude varies as a function of their absolute speed, the frames associated to these bodies cannot be perfectly inertial and they are not equivalent for the description of the physical laws. Therefore, the relativity principle cannot exactly apply. It is reduced to an approximation valid when the aether drift is weak, that is when $v/c \ll 1$. Note nevertheless that the laws of physics hardly vary when two frames $S_1$ and $S_2$ move with respect to one another at very low speeds $(v_{12}/c \ll 1)$.

(We should add that a principle of physics must be regarded as fundamental if it applies when the variables involved in the laws are exactly measured. If this is not the case, it loses its character of fundamental principle [1d]).

-In its original formulation of 1905 [6], unlike Poincaré's approach, Einstein's relativity principle dismisses the aether hypothesis. The postulate can be expressed as follows: "All the 'inertial frames' are equivalent. The laws of nature, including those of electromagnetism, take the same form in all of them". The arguments in favour of an aether drift developed in the previous and following chapters of the present text deprive the concept of relativity developed by Einstein of its unquestionable character and reduce it to an approximation valid when the aether drift is weak.

-Until now we have tested the consistency of special relativity in the field of dynamics as the laws are assumed, but without calling into question the compatibility of the relativity principle with mass-energy conservation and with the existence of mass, a question which will be tackled here.

Generally speaking, Galileo's relativity idea (as well as those of Poincaré and Einstein which derive from it) cannot be accepted without restrictions because it implies reciprocity, and, as we shall see, this is in contradiction with mass-energy conservation. Indeed, if the relativity principle did exactly apply in the physical world, the aether would exert an identical influence on all inertial frames* or no influence at all, and a perfect reciprocity between these frames would be observed. If this were not the case, the physical laws would prove different in two frames $S_0$ and S depending on the intensity of the aether drift. (This notion is implicit in the Einsteinian approach of the aether).

This kind of aether would offer no resistance to motion because if such a resistance did exist, it would be of different magnitude in the different frames, depending on their relative speed, in contradiction with the relativity principle.

The conventional concept of kinetic energy assumed by relativity is closely linked to the relativity principle and to the assumed absence of aether drift; indeed, insofar as there is no preferred frame, the kinetic energy has no absolute character. The following examples will put forward some of the paradoxes raised by this concept of kinetic energy.

Let us consider the case of a vehicle that travels from one inertial frame $S_1$ to another $S_2$. A part of its fuel provides the chemical energy k which is converted into kinetic energy. According to relativity, for an observer attached to frame $S_1$, when the vehicle reaches frame $S_2$, its kinetic energy has increased, but viewed by the observer attached to frame $S_2$, it has decreased. However, if chemical energy has been converted into kinetic energy during the travel, this must be true for all observers. This energy is not dependent on which one measures it. Chemical energy cannot give rise to a decrease of kinetic energy because, in such a case, the mass-energy conservation law would be transgressed.

---

* This is what Einstein means when in the conclusion of his book Ether and relativity he claims… this ether may not be thought of, as endowed with the quality characteristic of ponderable media, as consisting of parts which may be tracked through time. The idea of motion may not be applied to it. A Einstein, Sidelights on relativity, Dover, NY… Since the idea of motion may not be applied to this aether, there is no aether drift. This kind of aether does not offer resistance to motion.

(We must add that, for the observer standing in frame $S_2$, the part of chemical energy k mentioned above could not be converted into heat and exhaust energy, because heat and exhaust energy relate to the environment and not to the vehicle, and the two observers cannot draw opposite conclusions).

Such a paradox results from the fact that the kinetic energy of a body in relativity is regarded as strictly observer dependent. Indeed, the relativity principle requires that there is no preferred frame in which a body at rest has zero kinetic energy and from which the kinetic energy should be measured. If one assumes that the principle exactly applies in the physical world, a body is viewed as having zero kinetic energy by any observer at rest in the same frame as the body and, therefore, there is no storage of a well defined amount of kinetic energy when a body moves from one inertial frame to another.

There is no paradox any more if we consider that the total available kinetic energy is defined with respect to a privileged aether frame in which the absolute speed of any body at rest is zero. In this case, the total kinetic energy of a moving body has a well defined value and is not observer dependent. In the previous example, the increase of kinetic energy in the transfer from frame $S_1$ to frame $S_2$ would be absolute and recognized as the same by all observers. Conversely, in the transfer from $S_2$ to $S_1$, the decrease would also be recognized as the same by all. This implies that rest and motion are not only relative and that absolute speeds do exist.

We shall now present another example in which the paradoxes generated by the absence of aether drift and the existence of perfect inertial frames will be disclosed. To this end we shall make use of the criterion expressed in the first example which is required by logic, and we shall reason by contradiction. This topic which had already been tackled in ref [1], will be studied here in further detail.

Let us suppose that a spaceship leaves frame $S_1$ and, after acceleration, reaches a constant speed $v$ and becomes firmly attached to frame $S_2$. To this end, suppose that it has used an amount of fuel F capable of supplying the energy W, where W is the sum of the kinetic energy k and the heat and exhaust energy h, (W = k + h). We assume that the mass of the fuel is negligible compared to the mass of the spaceship. In $S_2$, the fuel tank is filled up again with a mass of fuel equal to F. If the frames $S_1$ and $S_2$ were perfectly inertial (equivalent), they would only be distinguished from one another by their relative speed $v$. This means that, if there was no aether drift, no difference could be observed in the physical properties of the transfers from $S_2$ to $S_1$ and from $S_1$ to $S_2$. Therefore, in order to come back to $S_1$, the spaceship should use the same amount of fuel as it does going from $S_1$ to $S_2$, a fact at the origin of the paradox. An observer standing in frame $S_2$ should note that the chemical energy used in the transit from $S_2$ to $S_1$ is W = k + h, the same as in the reverse direction. Assuming that $v/c \ll 1$, the observers in both frames should agree on that, and for the same reasons, they should agree on the fact that the amount of fuel which was to be converted into kinetic energy had to be the same in the two reverse directions. And this would be true no matter what point in $S_1$ the spaceship reaches upon its return. (Indeed, a body at rest with respect to a given inertial frame has a well-defined mass-energy whatever its position may be in this inertial frame. This mass-energy is equal to the sum of the internal mass-energy of the body $m_0C^2$ and its kinetic energy which is the same in any position in the inertial frame**).

Therefore, the spaceship would have used an amount of fuel equal to 2F corresponding to the energy 2k + 2h to leave and to recover its initial state in frame $S_1$. The heat and exhaust energy is conserved since it is released in the environment but, obviously, the kinetic energy is not, since with the assumed hypotheses a part of the fuel has been used to this end, while the final kinetic energy has not increased. And there would be no conservation of mass-energy. (Note that we have ignored the variation of mass with speed which, for $v \ll c$ is negligible. In any case, in this problem, it has no consequences on the conclusion drawn since, with the questionable hypotheses assumed, the amount of fuel used to be converted into kinetic energy is not null, while the final kinetic energy is obviously unchanged.)

Thus, mass-energy conservation, which is one of the most basic assumptions of physics, excludes the exact application of the relativity principle in the physical world. The implications of this paradoxical result will be developed below.

The point of view of the fundamental aether theory is completely different. Here there is no complete equivalence of frames $S_1$ and $S_2$. From $S_1$ to $S_2$ a part of the fuel must be converted into kinetic energy and from $S_2$ to $S_1$ the spaceship must restore this same energy to the environment as a form of heat. As a result, the mass-energy conservation law will be obeyed. The situation is similar to that of a body which acquires potential energy E when it moves from one level A, to another B. Upon its return, the body must restore the same energy E. But this implies that $S_1$ and $S_2$ experience a different influence from the substratum. We can conclude that the substratum offers resistance to motion, which increases with absolute velocity and which will be higher in $S_2$ than in $S_1$***. (Note that, in practice, the process we have just studied can be masked by the presence of an atmosphere and of strong gravitational forces).

---

** Of course this last statement would exactly apply only in an ideal inertial frame where there is no gravity and where the motion of the frame is strictly rectilinear and uniform.

Another consequence of the interaction of bodies with the aether is that the principle of inertia cannot rigorously apply (see later).

For the same reason, the total relativistic momentum of particles interacting in a collision cannot be exactly the same before and after the collision.

Furthermore, the relativity principle implies that the mass-energy available of a body A, at rest with respect to an inertial frame S, is not well-defined. It has only relative value with respect to another body B, and if the speed of body B tends towards the speed of light, the kinetic energy of A with respect to B will tend toward infinity. This is untenable. The total available mass-energy of a body is finite and is defined with respect to the fundamental frame. It is absurd to consider that it depends on the speed of another body.

We must conclude that the relativity principle is an abstract concept which cannot exactly apply in the physical world. It nevertheless remains that the (almost) uniform motion of a given reference frame is imperceptible for an observer at rest in this frame. But this does not result from the fact that only relative speeds have a meaning (and from the Galilean idea that motion is like nothing). Uniform absolute speed is not perceived, precisely because it remains unchanged, in other words, because the energy of the body in motion is not modified.

**On the origin of inertial mass**

As we have seen, the relativity principle implies that two identical bodies, attached to any two inertial systems $S_1$ and $S_2$, assume a completely

---

*** A body at rest in a frame S1 has a kinetic energy which is equal to the energy needed to overcome the aether resistance when the body moves from the aether frame S0 to S1. The principle of inertia would exactly apply only if, when the body is at rest in S1, the aether would no longer exert a pressure on the body, which is unrealistic knowing that the body has absolute kinetic energy and that the aether is responsible of length contraction and mass increase even when the body is at rest. Yet, this pressure must be weak enough (when v<<C) so that the principle of inertia applies almost exactly at low speeds. From S0 to S2 the kinetic energy acquired will be higher and then the aether pressure will also be higher.

symmetrical situation and therefore possess the same energy status.

If a spaceship needs to use fuel to move from $S_1$ to $S_2$, assuming that $v/c<<1$, it will also use the same amount of fuel to move from $S_2$ to $S_1$. If we suppose that the chemical energy used is h + k it would have used energy 2h + 2k on a round trip while, paradoxically, its final kinetic energy would not have increased and the energy will not be conserved. In order for the energy to be conserved, it should move from $S_1$ to $S_2$ (or from $S_2$ to $S_1$) without changing its kinetic energy. Such a paradoxical result can be easily explained when we know that in the absence of aether drift, there is no hierarchy between inertial frames.

Therefore, assuming that the rest mass of the spaceship is $m_0$ we should have:

$$(m - m_0)C^2 = 0$$
$$\Rightarrow m_0 C^2 \left[ \left(1 - v^2/C^2\right)^{-1/2} - 1 \right] = 0$$

where $m$ is the mass of the spaceship in $S_2$ viewed from $S_1$, and $v$ the speed of $S_2$ with respect to $S_1$. Since $v \neq 0$, the equation implies that $m_0 = 0$. So, paradoxically, the fact that the aether exerts no influence on the inertial frames would imply that the bodies do not possess mass.

On the contrary, the fundamental aether theory implies a different influence of the substratum on $S_1$ and $S_2$. Designating as $v_1$ the speed of $S_1$ with respect to the aether frame $S_0$ and $v_2$ the speed of $S_2$ with respect to $S_0$, the kinetic energy acquired by the spaceship when it moves from $S_1$ to $S_2$ will be:

$$m_0 C^2 \left[ \left(1 - v_2^2/C^2\right)^{-1/2} - \left(1 - v_1^2/C^2\right)^{-1/2} \right] \tag{29}$$

Upon its return, the part of chemical energy acquired by the spaceship as a form of kinetic energy, will be restored to the environment as a form of heat and therefore will be conserved. Thus, expression (29) is not null and $m_0 \neq 0$.

We can conclude that the existence of a mass $m_0 \neq 0$ depends not only on the quantity of matter, but also on the action of the aether on the physical bodies, which implies that the exact applicability of the relativity principle in the physical world is incompatible with the existence of mass.

**Mass-energy conservation**

This law must be viewed as unquestionable, because mass-energy cannot arises from nothing and, conversely, it cannot be destroyed. Nevertheless, in processes where mass or energy are exchanged, we must take account of the fact that the aether can absorb or supply part of the energy.

**Principle of inertia**

In a previous paper [35], we declared that any objection to the applicability of the relativity principle, also challenges the principle of inertia. It is important here to give further information in order to specify what we mean. In its original formulation, the principle was expressed in concrete terms, "a marble sliding on a perfectly smooth horizontal surface (without any friction) in vacuum, remains perpetually in its state of motion".

Of course, if, in agreement with the Galilean relativity principle, rest and uniform motion are only relative, we can consider that the marble is at rest in its reference system and, as a consequence, it must remain in this state of rest.

But, in the fundamental aether theory proposed here, absolute rest exists and is distinct from motion. The difference results from the existence of the aether. Under the action of the aether, the marble will experience a gradual slowing down, hardly perceptible, but not null. The Galilean principle of inertia is therefore called into question.

Now, in its modern sense, the principle of inertia can be expressed as: "a body not subjected to any physical force remains perpetually in its state of motion".

If we assume this more precise definition, the principle of inertia is essentially correct. Indeed, according to the theory proposed here, the aether should exert a pressure, very weak but which increases with absolute velocity, on the body, causing the body to slow down. If this pressure was balanced, the body would effectively remain in its state of motion. This condition is necessary in order for the law of mass-energy conservation to be obeyed.

**Speed of light invariance and Lorentz-FitzGerald's contraction**

This question has been extensively studied in ref [1a]. The new significant results published since the publication of this book will be briefly outlined here.

We are indebted to Builder and Prokhovnik [8], for having demonstrated that, assuming the anisotropy of the one-way speed of light and length contraction, the two-way transit time of light $2T$ along a rod attached to the Earth frame is isotropic. In other words, this time is independent of the angle separating the rod and the $x_0$, $x$ -axis of any co-ordinate system attached to the Earth frame. The value obtained by these authors was indeed:

$$2T = \frac{2\ell_0}{C\sqrt{1-v^2/C^2}} \tag{30}$$

Where $\ell_0$ is the length assumed by the rod when it is at rest in the aether frame, and *v* the speed of the Earth with respect to this privileged frame. In fact in a recent paper, Cahill and Kitto [27] claimed that perfect isotropy is observed only in vacuum. The Michelson-Morley experiments carried out in air or in gas allow detection of a fringe shift which until now has been considered as experimental error and then ignored. The magnitude of the aether drift deduced from the shift by the classical authors was of the order of 8 Km/sec in air. According to Cahill this result is inexact. The author declared "It is essential in analysing data, to correct for the refractive index effect". If the data from gas mode interferometers are analysed using Lorentz-FitzGerald's contraction the aether drift is found to be of the order of 400 Km/sec. (For further detail see the Entry relative to M. Allais in "Further references with comments")

It is too early to draw conclusion from the articles of these authors, but if the work is confirmed, it will provide another weighty argument in favour of the aether.

The demonstration carried out by Builder and Prokhovnik remains highly meaningful, yet, in the light of the new data, the two-way transit time of light along a rod would be perfectly isotropic only in the vacuum.

We must add that, contrary to what these authors asserted, *2T* is not the two-way transit time of light measured with clocks attached to the Earth frame since the measurement does not take account of clock retardation [1, 31]. Indeed, if *2T* was the value measured in the Earth frame, the *apparent* two-way speed of light in vacuum along the rod would not be found constant and of magnitude C.

Let us demonstrate this point. Notice first that when we measure the rod in a moving frame, (i, e, different from the aether frame) we do not find its real length. Since the meter stick used to measure the rod is also contracted, the length found is the rest length $\ell_0$ which the rod assumes in the aether frame.

Therefore, if Prokhovnik's claim were true, we would find for the speed of light in the Earth frame:

$$\frac{2\ell_0}{2\ell_0/C\sqrt{1-v^2/C^2}} = C\sqrt{1-v^2/C^2}$$

Since in reality, *2T* is the two-way transit time of light that clocks would display if they were attached to the aether frame, the measurement in the Earth frame gives $\frac{2\ell_0}{C}$ (from (30)) and therefore, the *apparent* two-way speed of light is found to be:

$$2\ell_0 \Big/ \frac{2\ell_0}{C} = C$$

This result also corresponds to the *apparent* one-way speed of light, since as demonstrated in [1] all the attempts to measure this velocity with clocks synchronized by means of the Einstein-Poincaré procedure, or by slow clock transport, only enable in fact to measure the *apparent* two-way speed of light.

**VIII. Conclusion**
Starting from the Galilean transformations and assuming the postulates of Lorentz, we have obtained a set of transformations applicable to any pair of 'inertial bodies' aligned along the direction of motion of the solar system. They take a different form from the Lorentz-Poincaré transformations.

In order to derive them, we were compelled to modify the Galilean transformations by taking account of the systematic measurement distortions. Conversely, they must be corrected in order to obtain the Galilean relations, which are the true transformations when no measurement distortions are present. (Of course this implies that when a body A moves at speed *v* with respect to a system of co-ordinates attached to the aether frame, the real speed relative to A of another body B moving along the direction OA will be limited to *v' < C − v*)

The extended space-time transformations derived in this text are equivalent to the inertial transformations derived by F. Selleri [11]; nevertheless, they take a different mathematical form, since the synchronization procedures utilized here are the usual procedures. F. Selleri makes use of the absolute synchronization procedure of Mansouri and Sexl [13] which is ideal, but should be really difficult to apply. In addition, we have demonstrated that the inertial transformations conceal hidden variables which are nothing else than the Galilean transformations ([see ref 1]).

The derivation is demonstrated to be consistent, since the extended transformations reduce to the Lorentz-Poincaré transformations when one of the frames they connect is the fundamental inertial frame [1d]. They also explain why the *apparent* (measured) velocity of light is found to be constant (although, after correction of the systematic measurement distortions, they show that the real one-way velocity of light is constant exclusively in the fundamental aether frame.)

These extended space-time transformations do contradict the application, in all generality, of the relativity principle in the physical world. Indeed, as we saw, while the principle can be used as an approximation when the absolute speed of bodies is very low in comparison to the speed of light, (in which case the space and time coordinates are almost exactly measured), it cannot be generalized. (See also [1, 35]). Moreover, as the principle *seems to apply*, even at high speed, provided that the space and time co-ordinates are altered by measurement distortions, it must be qualified as contingent [1d]. It cannot be considered as a fundamental principle of physics).

We must add that, insofar as the substratum interacts with the bodies or the particles present in a physical process, the laws of conservation would exactly apply only if this interaction was taken into account. At low speed with respect to the aether frame ($v \ll c$), the effect is certainly imperceptible, but this should not be the case when the particles move at speeds close to the speed of light.

As we have seen, the aether hypothesis enables us to account for several fundamental questions that had never found a satisfactory explanation before.

**Post Scriptum**

Although the present version provides additional explanations, the conclusions drawn do not differ from the previous version (physics/0604207, June 21 2006).

**Acknowledgements**

The author is grateful to Dr. Michael C. Duffy for giving him the opportunity of writing a chapter of this book and for the friendly relations he had with him in preparing this project. He is indebted to the late Pr. Jean Pierre Vigier for his unreserved support, to the late Pr. Victor Bashkov for his constant encouragement and to Pr. Franco Selleri for the approval given to some of his views. He would like to thank, Prs. William Cantrell, Max Jammer, Harold Puthoff and Ruggero Maria Santilli for the positive comments they have made on his manuscripts. He is also grateful to the late Pr. Simon Prokhovnik, with whom he has kept up a regular correspondence during more than two years.

**Further references with comments**
M. Allais, L'anisotropie de l'espace (Clement Juglar, Paris 1997) 750 pages.

The author comments on experiments he performed with a paraconic pendulum which, according to him, demonstrated the anisotropy of space. He also refers to the experiments of Miller and Esclangon, which lead to identical conclusions. Allais remarks:

Miller repeated the experiments of Michelson and Morley a number of times and concluded: "...Since the theory of relativity postulates an exact null effect from the aether drift experiment which had never been obtained in fact, the writer felt impelled to repeat the experiment in order to secure a definitive result." (L'anisotropie de l'espace, p 383).

Allais also notes (p. 405 of his book note 9 and p. 581 note 2) "it is erroneous to repeat that the experiments of Michelson and Morley of 1887 gave a completely negative result since they showed a fringe shift corresponding to a speed of 8 km/sec."

*Comments in the light of new data.*

The author does not quote the modern Michelson-Morley type experiments, the sensitivity of which has been considerably increased. In fact, the experiments performed in vacuum do not detect the aether drift. Vacuum interferometer experiments have always given almost null effects. This was the case for Joos (1930), and for Jaseja et al. (1964), who used masers mounted at right angles. Brillet and Hall in 1979 performed an experiment of very high sensitivity (about $30 \times 10^{-5}$) which showed a two-way transit time of light almost identical in two perpendicular directions. Confirmation has been given recently by Müller et al, Shiller et al, Herrmann et al, among others.

An attempt to reinterpret the results obtained by Miller has recently been made by Cahill and Kitto (see ref 27). Contrary to Brillet and Hall's experiment which was performed in vacuum, Miller's experiments were operated in gas mode. In their Apeiron 2003 article, Cahill and Kitto asserted that, after correcting for the refractive index effect of the air, Miller's experiments reveal an absolute speed of the Earth frame of $v= 335 \pm 57$ Km/sec, a result in agreement with the speed of $v=365 \pm 18$ km/sec determined from the dipole fit, in 1991, to the NASA COBE satellite Cosmic Background Radiation (CMB) observations. More recently Cahill has corrected the previous results asserting that the absolute motion of the solar system is a "vector sum of the universal

CMB velocity and the net velocity associated with the local gravitational inflows into the milky-way and the local cluster". The absolute speed was therefore corrected to yield $420 \pm 30$ Km/sec.

A. Brillet and J.L Hall, Phys. Rev. Lett 42 (1979) p 549.

According to Hayden it is "by far the best Michelson-Morley experiment performed to date. It has been designed to be clear in its interpretation and free of spurious effects".

The authors have handled their data in such a manner that effects that may arise from the Earth's rotation are ignored.

The experiment performed in vacuum could not detect the aether drift.

More recently the results obtained by Brillet and Hall have been confirmed and improved. See for example H. Müller et al, arXiv: physics/0305117, 28 May, 2003 and Phys. Rev. Lett, 91, 02040 (2003), S. Herrmann et al, arXiv: physics/0508097, 15 Aug 2005, S. Shiller et al, arXiv: physics/0510169 18 0ct 2005.

G. Builder, Aust. J. Phys 11 (1958a) 279 and 11 (1958b) p 457 and Philosophy Sci 26 (1959) p 135.

These articles are historically important; they develop an original viewpoint regarding relativity theory. The ideas of Builder have been presented and developed by his disciple Simon Prokhovnik in different articles and in two reference books. (See ref [8].)

M.C. Duffy, "Aether, cosmology and general relativity," and "The aether, quantum mechanics and models of matter," Gdansk conference, Sept. 1995. An extended version of the first article was published in the supplementary papers of the 1998 P.I.R.T. conference, Physical Interpretations of Relativity Theory, Imperial College London, 11-14 September 1998 p 16.

The first of these papers attempts a review of the relativistic world aether theories and seeks to identify the links between them. The aether concepts developed by Clube, Cavalleri, Borneas, Wheeler, Nesteruk, Prokhovnik, Ives, and Einstein, among others are analyzed.

The author concludes that a promising role for the aether within general relativity and cosmology has been convincingly demonstrated in recent years.

The second paper points out the role of the aether in the formulation of a grand comprehensive theory unifying relativity and quantum mechanics. References to Einstein, Dirac, Borneas, Podhala, Jennison, Winterberg, Cavalleri, Eddington, among others.

M.C. Duffy, "Science theory and reality" Lecture presented at the Meeting "Phenomenalism: Science, information and interpretations of reality", School of computing and advanced cybernetics, University of West England, Bristol 10$^{th}$ September 2005. This lecture develops and deepens the discussions tackled in the previous lectures and treats different subjects of epistemics such as "Realism and nominalism" the Einstein-Minskowski and Lorentz-Poincaré programmes are compared and the different concepts of aether are analysed.

L. Kostro, Physical interpretation of Albert Einstein's relativistic ether concept, Physical Interpretations of Relativity Theory (P.I.R.T.) 9-12 September 1994 p 206.

The author gives a thorough analysis of the evolution of Einstein's ideas about the aether.

"Until the end of his life, Einstein denied the existence of an ether as it was conceived in 19th century physics, in particular Lorentz's ether which was in the first place a privileged reference frame … because it violated his principle of relativity. Nevertheless, in 1916, Einstein proposed a new conception of the ether … which does not violate the principle of relativity because the space-time is conceived in it as a material medium sui-generis that can in no way constitute a frame of reference".

The author points out that in 1918 Lorentz also presented a model of ether at rest with respect to every reference frame, not only with respect to a preferred frame.

As we have seen, the model of ether we have presented in the previous chapters lends support to a privileged ether frame.

M. Mascart, Sur les modifications qu'éprouve la lumière par suite du mouvement de la source lumineuse et du mouvement de l'observateur (Annales scientifiques de l'Ecole Normale Supérieure) 2° serie, t I, p 157 -214, and p 364-420.

The author gives a detailed account of the results of a number of experiments designed to verify the influence of the Earth's motion on optical phenomena (experiments of Arago, Fresnel, Fizeau, measurement of the rotating power of quartz, double refraction, etc.)

The author concludes that the translation of the Earth has no influence on these optical phenomena. As a result, they do not detect the absolute motion of the Earth. Only relative motions can be observed.

The experiments analysed by Mascart have cleared the way for Potier and Veltmann. (See reference to these authors). However, the conclusions reached by Mascart cannot be extended to all types of experiments (for example, anisotropy of the CMB, muon flux anisotropy, Michelson experiment in gas mode, Marinov's experiments.) The conclusions of Mascart are limited to first order experiments.

T. E. Phipps Jr, Potier's principle, a trap for aetherists and others, Galilean Electrodynamics 3 (1992) p 56.

The author points out that Potier's principle denies the theoretical possibility of simple optical test of the existence of an aether wind (*to first order*).

Nevertheless as demonstrated in section II.3 (see also ref [1]) *second order aether wind* has been observed by different experimenters.

M. G. Sagnac, M.E. Bouty. L'ether lumineux démontré par l'effet du vent relatif d'ether dans un interferometre en rotation uniforme" and "Sur la preuve de la réalité de l'ether lumineux par l'expérience de l'interferographe tournant, Comptes-rendus Acad Sci Paris, vol 157, (1913) p 708 and 1410.

A non null fringe shift is observed when light is sent in opposite directions around a rotating table. The experiment lends support to the hypothesis of a non entrained aether.

R.M. Santilli, Confirmation of Don Borghi's experiment on the synthesis of neutrons from protons and electrons.
Arxiv physics/0608229, 23 Aug 2006
And Neutrino and/or etherino.
Arxiv physics/0610263, 28 0ct 2006, to be published.
The author comments on experiments performed by Don Carlo Borghi and his associates, who claimed having carried out the synthesis of the neutrons from protons and electrons in the late 1960s. He reports experimental and theoretical studies showing that under certain conditions, electric arcs, within a hydrogen gas, produce neutral, hadron size entities, he called pseudoneutrons which are absorbed by nuclei thus causing nuclear transmutations that confirm Don Borghi's experiment.

The author argues that the neutrino hypothesis is afflicted by a number of unresolved aspects. In particular the weak interaction with familiar reaction

$$p^+ + e^- \rightarrow n + \bar{\nu} \qquad (1)$$

violates the energy conservation law unless the proton and the electron have kinetic energy of at least 0.78 Mev, in which case no energy is left for the neutrino, a result due to the fact that the sum of the proton and the electron rest energies (0.78 Mev) is smaller than the neutron rest energy. Yet as the author argues, in the event that protons and electrons have a relative energy of at least 0.78 Mev, synthesis (1) is not permitted by quantum mechanics.

The alternative etherino hypothesis developed by the author enables conservation of energy via the interaction $p^+ + a_n + e^- \rightarrow n$ where $a_n$ refers to the *neutron etherino*, a particle absorbed from the ether having mass and charge 0, spin ½ and a minimum of 0.78 Mev energy. Several other arguments in the papers lend support to the assumed hypothesis.

Although our conception of the aether differs from the author's standpoint, his etherino hypothesis should be (with some adjustments) compatible with our concept of aether. Despite it questions some generally accepted concepts, the hypothesis deserves to be examined with complete impartiality.

F. Selleri. Lezioni di Relativita. (Ed Progedit, Bari, Italy, March 2003).

These lessons explain in an elementary but critical way the special theory of Relativity and the conceptual foundations of the general theory, giving ample place to the most important ideas and to their philosophical implications. In addition to the orthodox theory, the author presents the most important investigations made during the last ten years. According to him, there is an infinite number of theories equivalent to "special relativity" all based on the existence of a privileged reference frame. He asserts that certain phenomena break the equivalence, and are better explained using absolute synchronization. A return to Lorentz aether is finally possible.

Although our approaches are quite different, the ideas developed in this book show several points of convergence with our views. Nevertheless as we demonstrated in this text and in ref [1], if we assume the existence of a privileged reference frame and of an aether drift, the equivalence between special relativity and aether theories is reduced to an approximation valid when the absolute speed of bodies is weak (v<<c) (see in particular ref [1d]).

W. Veltmann, Astron Nachr vol.76 (1870) p 129-144, and A. Potier, Journal de physique (Paris) vol 3 (1874) p 201-204.

Using Fermat's principle the authors claim that it is impossible by means of an optical experiment to observe *a first order aether wind* in v/C. (However, it should be borne in mind that several modern experiments, described in section II.3, have detected *a second order aether wind*.)

C. K. Whitney "How can paradox happen"? in Physical interpretations of relativity theory (P.I.R.T), 15-18 September 2000, p 238.

Einstein relativity leaves us today with a number of paradoxes. The paper develops the view that physical reality is one thing while our conceptual model for it is quite another. When the two do not match, we will make strong inferences from data which can be inconsistent and lead to apparent paradoxes. The possibility that a wrong physical model may be embedded in Einstein's relativity theory is traced to the sequence of historical development: in the early days of his work, Einstein worked with Maxwell's electromagnetic theory but not modern quantum mechanics.

The proposed model includes facts that have appeared since the advent of quantum mechanics. The author asserts that it predicts the main features of special relativity without paradoxes as well as the main predictions of general relativity.